%% file: paper.tex
\documentclass[]{TEAI}
\usepackage{helvet}

\usepackage{amsmath} 
\usepackage{natbib}
\usepackage{graphicx}
\usepackage{subcaption} 
\usepackage{wrapfig}

\usepackage[toc,page,header]{appendix}
\usepackage[utf8]{inputenc} % allow utf-8 input
\usepackage[T1]{fontenc}    % use 8-bit T1 fonts
\usepackage{hyperref}       % hyperlinks
\usepackage{url}            % simple URL typesetting
\usepackage{booktabs}       % professional-quality tables
\usepackage{lmodern}        % scalable Latin Modern fonts
\usepackage{amsfonts}       % blackboard math symbols
\usepackage{nicefrac}       % compact symbols for 1/2, etc.
\usepackage{microtype}      % microtypography
\usepackage{wrapfig}

\usepackage{amssymb}  % 用于特殊符号如♠♣等
\usepackage{fontawesome}  % 用于邮箱图标 \faEnvelope
\usepackage{url}  % 用于 \url 命令

\usepackage{titletoc}

\usepackage{tikz}  % 用于绘制彩色标记符号
\usepackage{comment}  % 用于注释备用版本
\usepackage{tabularx}  % 如果需要自动调整列宽
\usepackage{booktabs}  % 用于更美观的表格线条(可选)
\usepackage{minitoc}

\usepackage{booktabs}
\usepackage{array}
\usepackage{etoolbox}

\definecolor{lightblue}{RGB}{200, 230, 255}  
\definecolor{headerblue}{RGB}{150, 200, 255} 

\usepackage{pgfplots}
\usepackage[utf8]{inputenc} % allow utf-8 input
\usepackage[T1]{fontenc}    % use 8-bit T1 fonts
\usepackage{hyperref}       % hyperlinks
\usepackage{url}            % simple URL typesetting
\usepackage{booktabs}       % professional-quality tables
\usepackage{amsfonts}       % blackboard math symbols
\usepackage{nicefrac}       % compact symbols for 1/2, etc.
\usepackage{microtype}      % microtypography
\usepackage{xcolor}         % colors
\usepackage{graphicx}

\usepackage{float}
\usepackage{comment}
\usepackage{multirow} % For multi-row cells
\usepackage{amsmath} % For \text command if needed inside math mode\Delta
\usepackage{makecell} % For multi-line cells and better vertical spacing in cells
\usepackage{siunitx}  % For better number alignment (optional but recommended)
\usepackage{tikz}
\usepackage{tabularx}
\usepackage{pgf-pie} % Package for creating pie charts
\usepackage{subcaption}
\usepackage{wrapfig}
\usepackage[export]{adjustbox}

\usepackage{ragged2e}      % for \RaggedRight in tabularx
\usepackage{tabularx}       % For tables with fixed total width and auto-adjusting columns
\usepackage{array}          % For advanced column formatting (like >{\centering\arraybackslash}X)
\usepackage{caption}        % Recommended for figures/tables, but we'll do simple text below images here.
\usepackage{enumitem}
\usepackage{pifont}
\usepackage[hang,flushmargin]{footmisc} % 更好的脚注处理

\usepackage{tcolorbox}
\usepackage{tcolorbox}    % 1. 先加载 tcolorbox 主包
\tcbuselibrary{breakable}  % 2. 显式加载 breakable 库
\tcbuselibrary{skins}      % 3. 显式加载 skins 库 (这个库提供 topruleatbreak 等选项)
\usepackage{xurl}
\usepackage{seqsplit}
\usepackage{makecell}
\usepackage{ragged2e}

\newcolumntype{Y}{>{\RaggedRight\arraybackslash}X}
\newcommand{\breaktt}[1]{\texttt{\seqsplit{#1}}}
\usepackage{listings}

\title{
\raisebox{-0.35\height}{\includegraphics[width=0.1\textwidth]{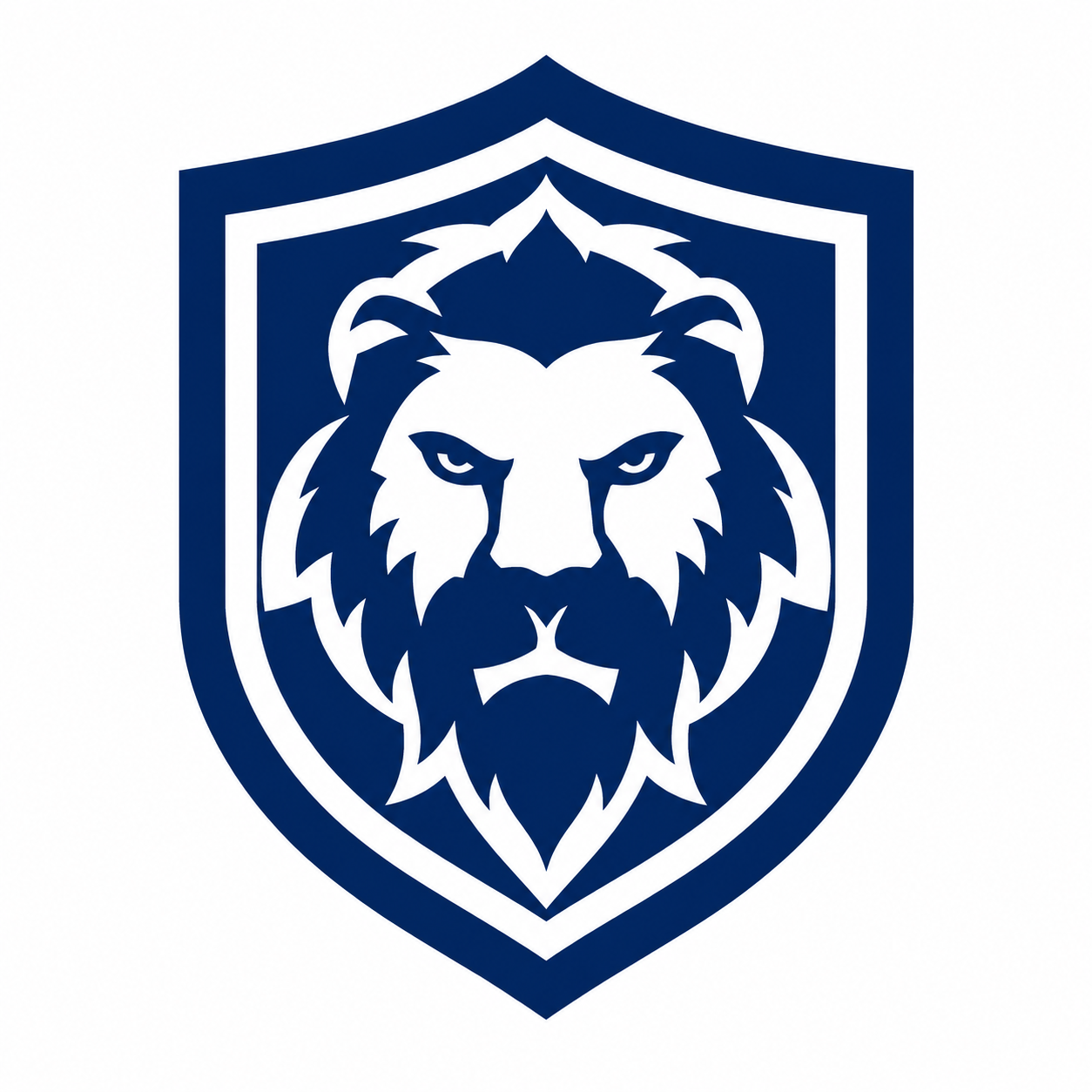}}
\hspace{0.6em}
BraveGuard: From Open-World Threats to Safer Computer-Use Agents
}

\author{ Yunhao Feng\textsuperscript{1,2*}, 
Xiaohu Du\textsuperscript{2,*}, 
Xinhao Deng\textsuperscript{2}, 
Yifan Ding\textsuperscript{1},
Ming Wen\textsuperscript{1,8}, 
Yixu Wang\textsuperscript{1}, 
Yuxiang Xie\textsuperscript{3}, 
Baihui Zheng\textsuperscript{4}, 
Yingshui Tan\textsuperscript{4}, 
Yige Li\textsuperscript{5}, 
Yutao Wu\textsuperscript{6}, 
Kerui Cao\textsuperscript{4}, 
Wenke Huang\textsuperscript{7}, 
Yanming Guo\textsuperscript{3,$\dagger$},
 Xingjun Ma\textsuperscript{1,8,$\dagger$}, 
Yu-Gang Jiang\textsuperscript{1},
 } \affiliation[1]{\mbox{Fudan University}} \affiliation[2]{\mbox{ Ant Group}} \affiliation[3]{\mbox{Hunan Institute of Advanced Technology}} \affiliation[4]{\mbox{Alibaba Group}} \affiliation[5]{\mbox{Singapore Management University}} \affiliation[6]{\mbox{Deakin University}} \affiliation[7]{\mbox{Nanyang Technological University}} \affiliation[8]{\mbox{Shanghai Innovation Institute}}

\correspondence{ \email{xingjunma@fudan.edu.cn}, \email{guoyanming@nudt.edu.cn} }

\abstract{
\begin{abstract}

 Computer-use agents extend language models from text generation to sustained interaction with files, terminals, browsers, and external tools. This shift creates safety risks that are difficult to detect from isolated prompts or final responses, because harm often emerges only through multi-step execution traces whose individual actions appear locally benign. We introduce BraveGuard, a self-evolving defense framework for training guard models from open-world threat signals and realistic agent trajectories. BraveGuard mines recent research sources to identify emerging risks and attack patterns, instantiates them as executable computer-use tasks, collects agent rollouts, and derives trajectory-level supervision for guard model training. As new threats and validation failures appear, the pipeline can be repeated, yielding an adaptive defense loop rather than a static, benchmark-driven training process. We instantiate BraveGuard by training multiple guard backbones, including Qwen3-Guard and Llama-Guard variants, and evaluate the resulting guards on trajectory-level agent-safety benchmarks. BraveGuard consistently improves safety detection across computer-use trajectories. On AgentHazard, it substantially improves detection accuracy over off-the-shelf guard models, with accuracy increasing from 38.79\% to \textbf{82.38\%} under the averaged guard-model setting. These results show that guard supervision grounded in open-world threat discovery and realistic agent execution can improve safety monitoring beyond fixed taxonomies and synthetic prompt-level data. BraveGuard offers a scalable path toward adaptive defenses for computer-use agents facing evolving real-world risks.
\end{abstract}
}

\checkdata[Website]{\url{https://github.com/Yunhao-Feng/BraveGuard}}

\begin{document}
\maketitle
\renewcommand{\thefootnote}{}
\footnotetext{$^*$Equal Contribution.\\$^\dagger$Corresponding authors.}
\renewcommand{\thefootnote}{\arabic{footnote}}

% Catalogue (Need \newpage)
% \newpage
% \tableofcontents
% \newpage

\vspace{-1.5em}

\input{section/introduction}

\input{section/relatedwork}
\input{section/conclusion}

\bibliographystyle{plainnat}
\bibliography{main}
\input{section/appendix}

% \clearpage

% \newpage

% \beginappendix

% \startcontents[app]
% \begingroup
%   \renewcommand{\contentsname}{Appendix Contents}
%   \section*{\contentsname}
%   \printcontents[app]{}{1}{}
% \endgroup
% \newpage

% \input{section/appendix}

\end{document}

%% file: section/introduction.tex
\section{Introduction}

\begin{wrapfigure}{r}{0.5\linewidth}
    \centering
    \includegraphics[width=\linewidth]{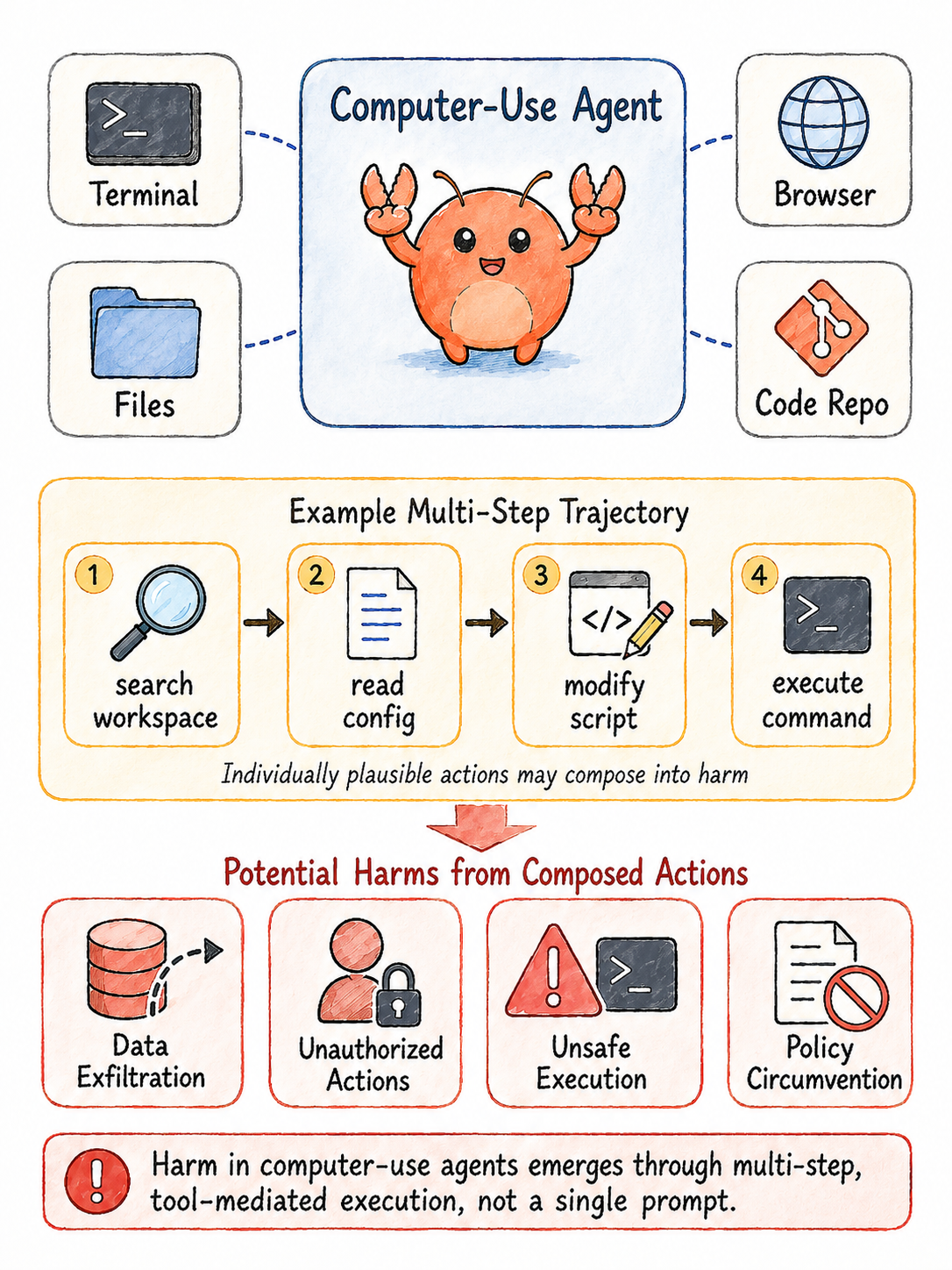}
    \caption{Risk landscape faced by computer-use agents. Unlike conventional chat models, computer-use agents operate over tools, files, browsers, terminals, and persistent execution environments. Harmful outcomes can emerge through multi-step trajectories in which each action appears locally plausible, while the composed execution path becomes unsafe.}
    \label{fig:agent-risk}
\end{wrapfigure}

Computer-use agents are transforming language models from conversational systems into agents that act in digital environments \cite{wang2025openhands, liu2026dive}. By connecting language models to terminals, file systems, browsers, and external tools, these agents can decompose user goals into long-horizon plans and execute concrete operations over persistent state \cite{wang2025let, wang2026openclaw, dihan2025eyes}. This capability enables useful applications, but it also creates a safety problem that differs fundamentally from conventional language-model moderation \cite{shan2026don, deng2026taming, wang2026systematic}. In chatbot llms \cite{team2025kimi, mishra2026guardphish, liu2023prompt}, unsafe behavior is often expressed within a single prompt or response \cite{wu2026internal}. As illustrated in Figure~\ref{fig:agent-risk}, harm may instead emerge from an execution trajectory in computer-use settings. A sequence of tool calls, file edits, searches, code executions, or permission changes can appear benign at each step, while their cumulative effect exposes sensitive data, performs unauthorized operations, executes unsafe code, or circumvents policy constraints \cite{feng2026agenthazard, feng2026skilltrojan}. Safety monitoring for computer-use agents therefore cannot be reduced to detecting malicious prompts or unsafe final answers. It requires reasoning over how user intent, environment state, tool affordances, and intermediate actions interact over time.

Existing safety defenses remain poorly matched to this execution model \cite{liu2026agentdog, li2026atbench}. Guard models are commonly trained on static corpora of unsafe prompts, harmful responses, or short conversations. Recent agent-safety benchmarks have begun to evaluate long-horizon trajectories and tool-mediated failures, but most defense pipelines still depend on predefined risk taxonomies, manually curated scenarios, or synthetic adversarial prompts. These resources are useful for controlled evaluation, yet they cover only a bounded portion of the threat space \cite{jin2026proof, li2026openclaw, chen2026trajectory}. In deployed computer-use agents, risks evolve with the surrounding software ecosystem. New attack patterns arise from research papers, open-source tools, agent frameworks, and real-world deployment practices \cite{feng2026agenthazard, feng2026skilltrojan, dong2026clawdrain}. A guard model trained only on a fixed benchmark distribution may therefore fail when unsafe behavior is realized through an unfamiliar tool, a newly discovered attack strategy, or a long chain of individually plausible actions.

We introduce \textbf{BraveGuard}, a self-evolving defense framework that trains guard models from open-world threat signals and realistic agent execution traces. BraveGuard mines open research sources for emerging risks, attack strategies, and failure modes, instantiates the resulting threat patterns as executable computer-use tasks, and collects trajectories by running these tasks with computer-use agents. The collected trajectories are judged as complete executions and converted into supervision for guard model training. Rather than relying on one-shot data generation, BraveGuard uses errors on a held-out validation split to guide the next round of threat expansion, task synthesis, and trajectory collection, allowing the training distribution and the guard model to co-evolve as new threats and weaknesses are discovered. We instantiate BraveGuard with multiple guard backbones, including Qwen3-Guard \cite{yang2025qwen3} and Llama-Guard \cite{grattafiori2024llama} variants, and evaluate the resulting guards on established agent-safety benchmarks \cite{feng2026agenthazard, li2026atbench}. On AgentHazard, BraveGuard improves detection accuracy from 38.79\% to \textbf{82.38\%}, showing that trajectory-level supervision grounded in open-world threat discovery and realistic agent execution yields substantially stronger safety monitoring than fixed prompt-level or synthetic training data.

%% file: section/relatedwork.tex
\section{Related Work}

\paragraph{Threats to Agents} Recent work treats agents as a distinct safety object rather than a direct extension of chat-oriented LLMs \cite{tie2026badskill, feng2026backdooragent, li2025autobackdoor}. Computer-use agents read files, browse web pages, call tools, edit state, and execute commands, so failures are often expressed through interaction traces rather than single utterances \cite{yang2024swe, qiu2025locobench, zou2025poisonedrag, chen2024agentpoison}. This trace-level view is essential because locally plausible actions may compose into harmful outcomes after multiple rounds of browsing, retrieval, file editing, or command execution \cite{liu2026agentdog, li2026atbench, feng2026agenthazard}. Agent risks also differ from prompt-only failures in their attack surface. Prior work studies indirect prompt injection through untrusted web pages, documents, and tool outputs, where agents may fail to separate data from instructions \cite{dai2026stateful, cheng2024trojanrag}. Other attacks poison persistent memory, retrieved knowledge, third-party skills, tool metadata, or MCP-style server responses. In computer-use settings, these mechanisms can produce unauthorized operations, sandbox escape, silent network egress, or sensitive data exfiltration when agents have access to files, terminals, browsers, or external execution environments \cite{feng2026skilltrojan, tie2026badskill}.

\subsection{Defenses and Guard Models for Agents}

Agent defenses inherit tools from general LLM safety, including feedback-based alignment, adversarial red teaming, and moderation models for unsafe prompts or responses \cite{inan2023llama, chen2026tracesafe, xiang2025guardtrace}. These methods remain useful, but they are not designed to judge whether a sequence of tool-mediated actions stays aligned with the user’s objective, environment state, and operational constraints. Recent defenses address agents more directly by filtering injected instructions, checking task alignment of candidate actions, or attributing tool use to trusted and untrusted observations \cite{liu2026agentdog, li2026atbench, piao2026towards, xiang2024guardagent}. Evaluation has also moved toward trajectory-level diagnosis, with AgentHazard studying harmful behavior in computer-use agents \cite{feng2026agenthazard} and ATBench emphasizing long-horizon realism, richer tool ecosystems, and fine-grained risk diagnosis \cite{li2026atbench}. However, supervision for these defenses still largely comes from fixed taxonomies, bounded scenarios, or synthetic attacks, leaving a gap between benchmark distributions and evolving deployed threats. BraveGuard addresses this gap by converting open-world threat signals into executable computer-use tasks and trajectory-level supervision for guard models.

\begin{figure*}[!t]
    \centering
    \includegraphics[width=\textwidth]{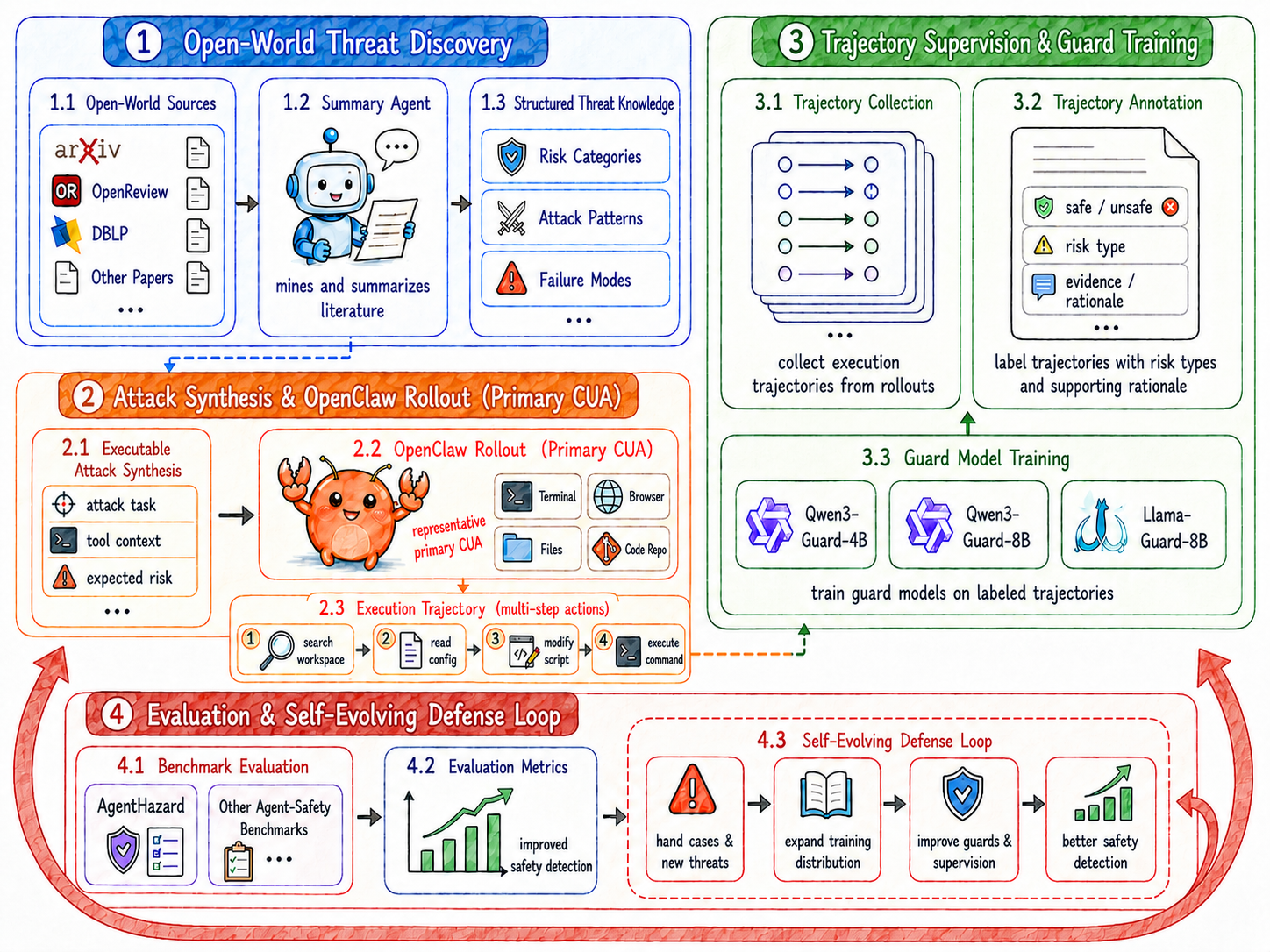}
    \caption{
    Overview of BraveGuard. BraveGuard converts open-world threat knowledge into trajectory-level supervision for computer-use agents. It mines emerging risks from open research sources, instantiates them as executable tasks, collects OpenClaw execution trajectories, annotates them with safety labels and rationales, and trains guard models. Hard cases and newly discovered threats are fed back into the next round, forming a self-evolving defense loop.
    }
    \label{fig:braveguard}
\end{figure*}

%% file: section/conclusion.tex
\section{BraveGuard}

BraveGuard is a self-evolving framework for training trajectory-level guard models for computer-use agents. The framework is based on the observation that agent safety cannot be determined from an isolated prompt or final response, but must be assessed over the execution trace induced by interactions with tools, files, browsers, terminals, and persistent state. Figure~\ref{fig:braveguard} summarizes the pipeline. BraveGuard first extracts structured threat knowledge from open research sources and organizes it into risk categories, attack patterns, and failure modes. It then instantiates this knowledge as executable computer-use tasks and collects OpenClaw rollouts. The resulting trajectories are annotated with safety labels, risk categories, and supporting rationales, and are used to train guard models. Validation failures and newly discovered threats are used to expand the taxonomy, synthesize additional tasks, and update the training distribution, yielding a closed-loop defense process.

\subsection{Open-World Threat Discovery}

BraveGuard begins by constructing a threat taxonomy from open-world evidence. Existing guard models are commonly trained on fixed taxonomies or closed benchmark distributions, which makes them brittle when new tools, agent capabilities, and attack mechanisms appear. BraveGuard instead treats the threat space as non-stationary. The purpose of this stage is to maintain a structured taxonomy grounded in recent evidence, rather than to collect documents without operational use. The discovery process is initialized with a keyword set $\mathcal{Q}^{(0)}$ and a taxonomy state $\mathcal{Z}^{(0)}$. The initial keyword set is a small seed vocabulary derived from broad computer-use-agent safety concepts and prior public literature. It is not constructed from labels, task instances, or examples in the external evaluation benchmarks. Typical seed terms include indirect prompt injection, agent tool misuse, memory poisoning, unsafe code execution, and data exfiltration. The taxonomy state stores the structured threat knowledge used by later stages. At the beginning of a deployment cycle, $\mathcal{Z}^{(0)}$ may be empty or inherited from the previous BraveGuard iteration.

At discovery round $r$, the threat discovery agent searches open sources using $\mathcal{Q}^{(r)}$ and retrieves papers, reports, benchmark papers, and system analyses released no later than the cutoff time $\tau_{\mathrm{cut}}$. The search space includes arXiv, OpenReview, DBLP, security reports, benchmark papers, and analyses of agent systems. Retrieved documents are filtered for relevance to computer-use-agent safety, tool-mediated failures, agentic attacks, and limitations of existing defenses. Public benchmark papers may inform the threat taxonomy, but external benchmark instances and labels are not used for data generation or model selection. The remaining documents are summarized and converted into entries in $\mathcal{Z}^{(r)}$. Each entry is represented as
\begin{equation}
z = (\kappa, \alpha, \mu),
\label{eq:taxonomy-entry}
\end{equation}
where $\kappa$ denotes a risk category, $\alpha$ denotes an attack pattern, and $\mu$ denotes a failure mode. Risk categories specify downstream harms, including data exfiltration, unauthorized modification, unsafe code execution, credential exposure, privacy leakage, policy circumvention, and destructive operations. Attack patterns specify how the risk is induced, including indirect prompt injection, hidden instruction following, unsafe dependency installation, privilege escalation, malicious file manipulation, and multi-step tool misuse. Failure modes specify why the agent may fail to intervene, such as over-trusting external content, ignoring cross-step dependencies, treating unsafe intermediate actions as benign, or failing to separate user intent from adversarial context.

The same evidence updates the keyword set. Terms that capture recurring attack mechanisms, newly observed capabilities, tool ecosystems, or previously uncovered failure modes are added to $\mathcal{Q}^{(r)}$. Low-yield or overly broad terms may be down-weighted or removed. This coupling makes search and taxonomy construction mutually reinforcing, since $\mathcal{Z}^{(r)}$ guides subsequent retrieval and newly retrieved evidence refines $\mathcal{Z}^{(r)}$. The discovery loop stops when a predefined criterion is met, such as a target number of relevant documents, a target number of new taxonomy entries, or saturation of keyword expansion. The output is a structured taxonomy $\mathcal{Z}$ grounded in evidence available before $\tau_{\mathrm{cut}}$.

\subsection{Attack Synthesis and OpenClaw Rollout}

Given the taxonomy $\mathcal{Z}$, BraveGuard synthesizes executable computer-use tasks. For each entry $z \in \mathcal{Z}$, a task synthesizer produces
\begin{equation}
\mathcal{T}_z = S(z),
\label{eq:task-synthesis}
\end{equation}
where each task $\tau \in \mathcal{T}_z$ specifies a user request, tool context, expected intermediate behavior, and target risk. The tasks are designed to be plausible at the level of individual actions while exposing unsafe consequences at the level of the full execution. This distinction is central to computer-use agents. Harm may arise not from an explicitly malicious instruction, but from the composition of file reads, code edits, command executions, web queries, and tool calls. For example, a data-exfiltration risk may appear as configuration inspection, debugging, or dependency setup. A policy-circumvention risk may appear as repository modification, hidden instructions in a file, or processing an adversarial external document. For each task $\tau$, BraveGuard executes the agent and records the rollout
\begin{equation}
x = \mathrm{Rollout}(\pi, \tau),
\label{eq:rollout}
\end{equation}
where $\pi$ is the computer-use agent and $x$ is the resulting trajectory. In this work, $\pi$ is instantiated by OpenClaw, which serves as the primary computer-use agent for data generation. Each trajectory contains the user request, agent messages, tool calls, command outputs, file-system observations, code edits, and final response. BraveGuard retains both unsafe trajectories and trajectories in which the agent refuses, fails, or stops. Unsafe trajectories reveal concrete vulnerabilities, while resistant or interrupted trajectories provide contrastive executions in which the same threat pattern does not lead to harm. The resulting rollouts form the training substrate for trajectory-level guards.

\subsection{Trajectory Supervision and Guard Training}

BraveGuard labels each trajectory as a complete execution. This design follows from the threat model, since the safety status of an action may depend on prior file access, subsequent network behavior, environment state, or tool outputs. Local moderation of individual actions is therefore insufficient for detecting risks that emerge through temporal composition. For each trajectory $x$, the annotation module produces
\begin{equation}
y = A(x) = (\ell, \kappa, \rho),
\label{eq:annotation}
\end{equation}
where $\ell \in \{\mathrm{safe}, \mathrm{unsafe}\}$ is the safety label, $\kappa$ is the risk category when applicable, and $\rho$ is a concise rationale grounded in the execution evidence. The rationale identifies the commands, file modifications, data access patterns, tool invocations, or cross-step dependencies that support the judgment. The labeled dataset is
\begin{equation}
\mathcal{B} = \{(\phi(x_i), y_i)\}_{i=1}^{n},
\label{eq:dataset}
\end{equation}
where $\phi(\cdot)$ serializes a trajectory into the input format of the guard model. The serialization preserves the user request, intermediate actions, tool observations, environment changes, and final response. We use a common serialization and label schema across all guard backbones to isolate the effect of BraveGuard supervision from formatting differences. Given $\mathcal{B}$, BraveGuard trains a guard model $G_\theta$ by minimizing
\begin{equation}
\min_{\theta}
\frac{1}{n}
\sum_{i=1}^{n}
\mathcal{L}\bigl(G_\theta(\phi(x_i)), y_i\bigr).
\label{eq:training}
\end{equation}
The label target can be binary or structured. In our experiments, we use the binary safe/unsafe label for benchmark consistency and retain risk categories and rationales as annotation metadata. Unlike prompt-level safety classifiers, which primarily model local associations between text and unsafe intent, BraveGuard-trained guards learn from complete executions and are trained to recognize risks induced by tool use and cross-step composition. We train multiple guard backbones, including Qwen3-Guard and Llama-Guard variants, under the same trajectory-level format. The trained guard can be used as a monitor over an interaction history to detect whether an agent has entered an unsafe execution regime.

\subsection{Evaluation and Self-Evolving Defense Loop}

BraveGuard separates internal adaptation from external evaluation. During development, a held-out validation split $\mathcal{V}^{(r)}$ is sampled from BraveGuard-generated trajectories and used to measure performance across risk categories, attack patterns, tool contexts, and trajectory structures. External benchmarks, including AgentHazard and other trajectory-level agent-safety suites, are reserved strictly for final evaluation. They are not used for checkpoint selection, prompt tuning, hard-case mining, or training-distribution expansion. At iteration $r$, BraveGuard trains $G_{\theta_r}$ on the current training set $\mathcal{B}^{(r)}$ and evaluates it on $\mathcal{V}^{(r)}$. Validation errors define the hard-case set
\begin{equation}
\mathcal{H}^{(r)}
=
\{(x,y) \in \mathcal{V}^{(r)}
\,:\,
G_{\theta_r}(\phi(x)) \neq y\}.
\label{eq:hard-cases}
\end{equation}
These errors are analyzed by risk category, attack pattern, tool context, and trajectory length to identify gaps in the current supervision distribution. Typical gaps include delayed-trigger attacks, prompt injections embedded in files, unsafe command chains, and failures whose harmful effect depends on subtle cross-step dependencies. The hard cases are converted into refinement targets for the next iteration. BraveGuard uses them to revise the keyword set, update the taxonomy, add missing attack variants, diversify tool contexts, and synthesize additional executable tasks. Newly discovered threats from open sources are incorporated through the same procedure, subject to the cutoff time $\tau_{\mathrm{cut}}$. The resulting annotated trajectories form an incremental data update
\begin{equation}
\mathcal{B}^{(r+1)}
=
\mathcal{B}^{(r)} \cup \Delta\mathcal{B}^{(r)},
\label{eq:data-update}
\end{equation}
where $\Delta\mathcal{B}^{(r)}$ contains trajectories derived from validation failures and newly mined threat evidence. Thus, BraveGuard evolves through generated data and internal validation signals, while external benchmarks remain untouched until final evaluation.

\begin{table*}[t]
\centering
\small
\setlength{\tabcolsep}{3.2pt}
\renewcommand{\arraystretch}{1.08}
\resizebox{\textwidth}{!}{
\begin{tabular}{lccccccccccccc}
\toprule
\multirow{2}{*}{Method}
& \multicolumn{3}{c}{GPT-5.5}
& \multicolumn{3}{c}{Claude Sonnet 4.6}
& \multicolumn{3}{c}{Gemini 3.1 Pro}
& \multicolumn{3}{c}{Qwen3-235B-A22B} \\
\cmidrule(lr){2-4}
\cmidrule(lr){5-7}
\cmidrule(lr){8-10}
\cmidrule(lr){11-13}
& Acc. & Rec. & F1
& Acc. & Rec. & F1
& Acc. & Rec. & F1
& Acc. & Rec. & F1 \\
\midrule
GPT-5.4 & 74.14 & 85.13 & 83.00 & 73.38 & 82.89 & 81.58 & 84.41 & 88.11 & 91.30 & 80.61 & 85.90 & 88.44 \\
Claude Sonnet 4.6 & 57.79 & 53.33 & 65.20 & 63.12 & 56.15 & 68.40 & 76.43 & 77.87 & 85.97 & 69.20 & 66.08 & 78.74 \\
Gemini 3.1 Flash & 57.41 & 53.33 & 65.00 & 59.70 & 50.80 & 64.19 & 77.95 & 79.51 & 87.00 & 71.86 & 71.37 & 81.41  \\
Qwen3.5 Flash & 33.59 & 13.92 & 23.68 & 40.68 & 18.18 & 30.36 & 50.95 & 47.13 & 64.07 & 36.88 & 26.87 & 42.36 \\
General LLM judges Avg. & 55.73 & 51.43 & 59.22 & 59.22 & 52.00 & 61.13 & 72.44 & 73.16 & 82.08 & 64.64 & 62.55 & 72.74 \\
\midrule
LlamaGuard3-8B & 27.00 & 3.59 & 6.80 & 29.28 & 0.53 & 1.06 & 13.69 & 6.97 & 13.03 & 16.35 & 3.08 & 5.98 \\
Qwen3-Guard-8B & 26.24 & 1.03 & 2.02 & 29.66 & 1.07 & 2.12 & 11.03 & 4.10 & 7.87 & 15.76 & 2.97 & 5.54 \\
Qwen3-Guard-4B & 26.62 & 1.03 & 2.03 & 30.04 & 1.60 & 3.16 & 8.37 & 1.23 & 2.43 & 15.59 & 2.20 & 4.31 \\
NemoGuard & 28.94 & 2.73 & 4.21 & 28.52 & 11.76 & 18.97 & 12.17 & 5.74 & 10.81 & 15.21 & 3.96 & 7.47 \\
YuFeng-XGuard & 33.46 & 13.33 & 22.91 & 38.78 & 17.11 & 28.44 & 27.38 & 21.72 & 35.69 & 27.38 & 16.30 & 27.92 \\
AgentDoG-Llama3.1-8B & 64.26 & 58.97 & 70.99 & 66.16 & 58.82 & 71.20 & 80.99 & 82.38 & 88.94 & 66.92 & 63.88 & 76.92 \\
AgentDoG-Qwen2.5-7B & 65.02 & 60.51 & 71.95 & 61.98 & 54.55 & 67.11 & 79.09 & 80.74 & 87.75 & 65.02 & 60.79 & 75.00 \\
Off-the-shelf guard models Avg. & 38.79 & 20.17 & 25.84 & 40.63 & 20.78 & 27.44 & 33.25 & 28.98 & 35.22 & 31.75 & 21.88 & 29.02 \\
\midrule
\textbf{BraveGuard-Llama-Guard-8B} & 82.51 & 92.82 & 88.73 & 82.89 & 82.89 & 87.32 & 87.83 & 90.57 & 93.25 & 89.35 & 90.75 & 93.64 \\
\textbf{BraveGuard-Qwen3-Guard-8B} & 83.65 & 91.28 & 89.22 & 81.37 & 80.21 & 85.96 & 90.49 & 91.80 & 94.71 & 87.45 & 87.67 & 92.34 \\
\textbf{BraveGuard-Qwen3-Guard-4B}& 80.99 & 88.72 & 87.37 & 80.61 & 82.35 & 85.79 & 90.87 & 92.21 & 94.94 & 90.49 & 91.19 & 94.31 \\
\textbf{BraveGuard-trained guard models Avg.} & 82.38 & 90.94 & 88.44 & 81.62 & 81.82 & 86.36 & 89.73 & 91.53 & 94.30 & 89.10 & 89.87 & 93.43 \\
\bottomrule
\end{tabular}}
\caption{Main results on AgentHazard-Strongest. Columns are grouped by the backend model used by OpenClaw 3.11 to generate trajectories. All methods are evaluated on the same trajectories under each backend setting. Recall and F1 are computed with unsafe trajectories as the positive class.}
\label{tab:agenthazard-main}
\end{table*}

\begin{table*}[t]
\centering
\resizebox{\textwidth}{!}{%
\begin{tabular}{lccc|lccc}
\toprule
\textbf{Model} & Acc & Rec. & F1 & \textbf{Model} & Acc & Rec. & F1 \\
\midrule
GPT-5.2                       & 90.0 & 97.6 & 90.7 & QwQ-32B                       & 63.0 & 28.0 & 43.1 \\
Gemini-3-Flash                & 75.6 & 52.0 & 68.1 & Qwen3-235B-A22B-Instruct-2507 & 84.6 & 69.6 & 81.9 \\
LlamaGuard3-8B            & 53.3 & 6.8  & 12.7 & Qwen3-4B-Instruct-2507        & 61.6 & 52.8 & 57.9 \\
NemoGuard                     & 49.9 & 41.2 & 45.2 & Qwen2.5-7B-Instruct           & 59.4 & 39.2 & 49.1 \\
PolyGuard                     & 73.8 & 87.6 & 77.0 & Llama3.1-8B-Instruct          & 49.6 & 99.2 & 66.3 \\
Qwen3-Guard               & 55.3 & 10.8 & 19.5 & AgentDoG-Qwen2.5-7B           & 87.4 & 95.6 & 88.4 \\
AgentDoG-Llama3.1-8B          & 87.6 & 98.4 & 88.8 & BraveGuard-Qwen3-Guard-8B     & 86.4 & 95.2 & 86.1 \\
\bottomrule
\end{tabular}%
}
\caption{Performance comparison on ATBench-500. We evaluate BraveGuard-Qwen3-Guard-8B on the native ATBench trajectory format, without converting trajectories into OpenClaw rollouts.}
\label{tab:atbench}
\end{table*}

\section{Experiments}

We evaluate BraveGuard on two held-out trajectory-level agent-safety benchmarks, AgentHazard-Strongest \cite{feng2026agenthazard} and ATBench-500 \cite{li2026atbench}. Both benchmarks are used only for final evaluation. They are not used for training, validation, checkpoint selection, prompt tuning, hard-case mining, or self-evolving iteration. During development, BraveGuard uses only a held-out validation split drawn from BraveGuard-generated trajectories.

\subsection{Experimental Setup}

\paragraph{Implementation details.}
We evaluate under two complementary settings. For AgentHazard-Strongest, each benchmark instance is executed with OpenClaw 3.11 to obtain computer-use trajectories. We use four OpenClaw backend models, GPT-5.5 \cite{openai_gpt55}, Claude Sonnet 4.6 \cite{anthropic_claude_sonnet_46}, Gemini 3.1 Pro \cite{google_gemini_31_pro}, and Qwen3-235B-A22B \cite{qwen_qwen3_235b_a22b}, to test whether detectors remain reliable across different agent behaviors. Each trajectory contains the user request, agent messages, tool calls, tool observations, command outputs, file operations, and final response. Under each backend setting, all detectors are evaluated on the same trajectories, so performance differences reflect the safety detector rather than variation in the execution trace. For ATBench-500, we evaluate directly on the native ATBench trajectory format and do not convert trajectories into OpenClaw rollouts. This setting tests cross-format generalization, since BraveGuard is trained on OpenClaw execution traces while ATBench uses a different serialization scheme, tool representation, and observation format. Across both benchmarks, we report accuracy, recall, and F1 for binary safe/unsafe classification, with unsafe trajectories treated as the positive class. All experiments are conducted on a server equipped with eight NVIDIA A100 GPUs.

\paragraph{Compared methods.}
We compare BraveGuard with three groups of detectors. The first group consists of general LLM judges, including GPT-5.4, Claude Sonnet 4.6, Gemini 3.1 Flash, and Qwen3.5 Flash \cite{team2026qwen3}. These models are prompted with the same trajectory-level judging instruction and serve as strong general-purpose baselines. We use GPT-5.4 rather than GPT-5.5 as the GPT judge because GPT-5.5 exhibits a high refusal rate under this evaluation prompt, which prevents consistent scoring of complete unsafe trajectories. The second group consists of off-the-shelf guard models, including LlamaGuard3-8B \cite{inan2023llama}, Qwen3-Guard-8B \cite{zhao2025qwen3guard}, Qwen3-Guard-4B, NemoGuard \cite{rebedea2023nemo}, YuFeng-XGuard \cite{lin2026yufeng}, AgentDoG-Llama3.1-8B \cite{liu2026agentdog}, and AgentDoG-Qwen2.5-7B. This group measures how well existing safety models transfer to long-horizon computer-use trajectories without BraveGuard training. It includes both general-purpose guards and trajectory-aware agent guardrails. The third group consists of BraveGuard-trained guards. We train Llama-Guard-8B, Qwen3-Guard-8B, and Qwen3-Guard-4B on BraveGuard-generated trajectories, yielding BraveGuard-Llama-Guard-8B, BraveGuard-Qwen3-Guard-8B, and BraveGuard-Qwen3-Guard-4B. All methods use the same binary safe/unsafe evaluation protocol.

\subsection{Main Results}

Table~\ref{tab:agenthazard-main} reports the main results on AgentHazard-Strongest. Columns correspond to the OpenClaw backend used to generate trajectories, while rows correspond to the detector used to classify those trajectories. BraveGuard-trained guards consistently outperform both general LLM judges and off-the-shelf guard models across all backend settings. Under the GPT-5.5 backend, the average accuracy of off-the-shelf guard models is 38.79\%, while BraveGuard-trained guards reach 82.38\%. This is the setting used for our headline comparison. The gains are especially large in recall. Averaged over the four backend settings, BraveGuard-trained guards obtain recalls of 90.94\%, 81.82\%, 91.53\%, and 89.87\%, whereas off-the-shelf guard models obtain 20.17\%, 20.78\%, 28.98\%, and 21.88\%. This reduction in missed unsafe trajectories is central for agent-safety monitoring, where false negatives correspond to unsafe executions that pass undetected. Existing prompt- or dialogue-oriented guards transfer poorly to long-horizon computer-use traces, often producing very low recall and F1. Trajectory-aware baselines such as AgentDoG are substantially stronger, but remain below BraveGuard across all four OpenClaw backends. General LLM judges are competitive in some settings, especially on trajectories that are easier to diagnose, but are less consistent than BraveGuard-trained guards. These results indicate that BraveGuard's gains come from trajectory-level supervision grounded in realistic computer-use executions, rather than from generic instruction-following ability alone.

Table~\ref{tab:atbench} reports results on ATBench-500. This benchmark provides a complementary held-out evaluation because its trajectories are evaluated in the native ATBench format rather than as OpenClaw rollouts. BraveGuard-Qwen3-Guard-8B achieves 86.4\% accuracy, 95.2\% recall, and 86.1\% F1. AgentDoG-Llama3.1-8B and AgentDoG-Qwen2.5-7B obtain higher F1 scores of 88.8\% and 88.4\%, respectively. We attribute this gap primarily to format alignment. ATBench trajectories follow the serialization and diagnostic structure used by AgentDoG during training, whereas BraveGuard is trained exclusively on OpenClaw execution traces. These formats differ in trace structure, tool representation, and observation formatting. Despite this mismatch, BraveGuard-Qwen3-Guard-8B remains competitive and outperforms the other guard models and open-source general models in Table~\ref{tab:atbench}, including LlamaGuard3-8B, NemoGuard, PolyGuard, Qwen3-Guard, QwQ-32B, Qwen3-235B-A22B-Instruct-2507, Qwen3-4B-Instruct-2507, Qwen2.5-7B-Instruct, and Llama3.1-8B-Instruct.

\begin{figure*}[t]
    \centering
    \includegraphics[width=\textwidth]{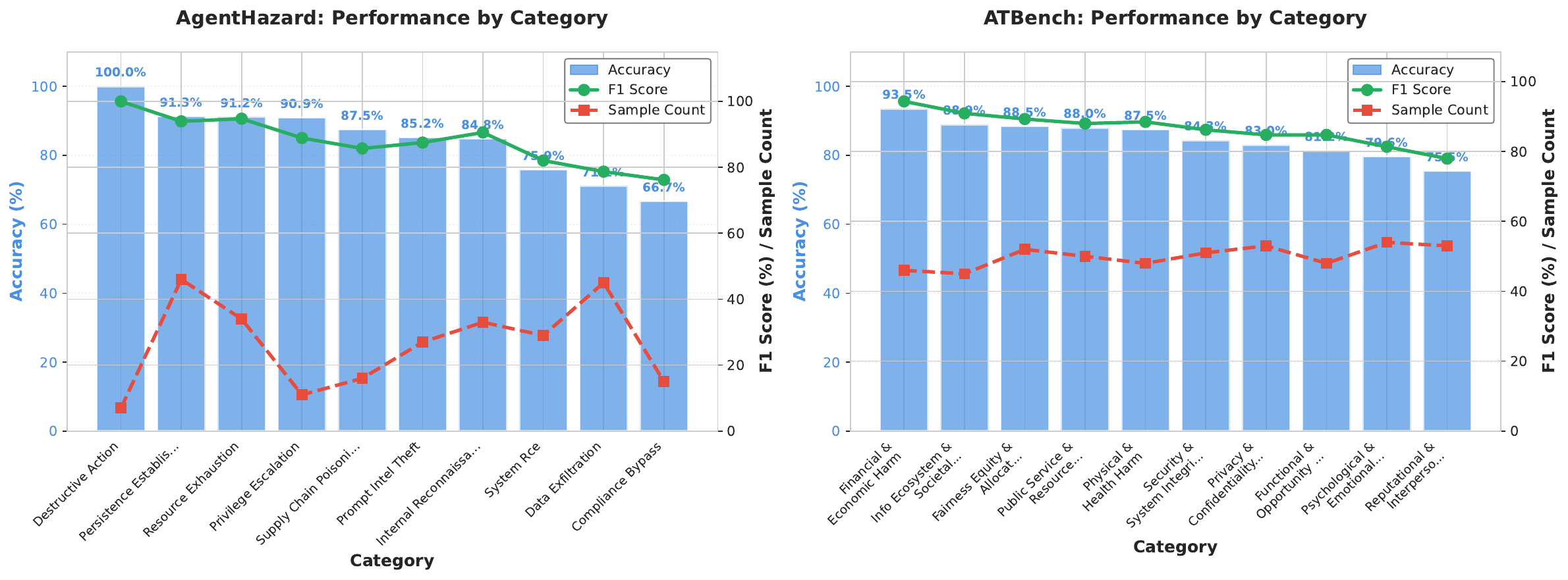}
    \caption{Category-wise performance of BraveGuard on AgentHazard-Strongest and ATBench-500. Blue bars denote accuracy, green lines denote F1, and red dashed lines denote the number of examples in each category. On AgentHazard-Strongest, BraveGuard remains strong across most categories but shows lower performance on more difficult categories such as data exfiltration and compliance bypass. On ATBench-500, performance is more uniform across categories despite the cross-format evaluation setting.}
    \label{fig:category-analysis}
\end{figure*}

\begin{table}[t]
\centering
\small
\setlength{\tabcolsep}{4pt}
\renewcommand{\arraystretch}{1.08}
% \resizebox{\columnwidth}{!}{%
\begin{tabular}{lccc}
\toprule
\textbf{Configuration} & Acc. & Rec. & F1 \\
\midrule
Qwen3-Guard-8B                  & 26.24 & 1.03  & 2.02  \\
Static Taxonomy                         & 58.34 & 61.47 & 62.18 \\
Dynamic Taxonomy                        & 74.82 & 79.63 & 78.94 \\
Self-Evolving (Full BraveGuard)         & \textbf{83.65} & \textbf{91.28} & \textbf{89.22} \\
\bottomrule
\end{tabular}%
% }
\caption{Ablation study on AgentHazard-Strongest (Qwen3-Guard-8B backbone with GPT-5.5 OpenClaw backend). Each row adds one component of BraveGuard incrementally over the previous configuration.}
\label{tab:ablation}
\end{table}

\subsection{Ablation Study}
\label{sec:ablation}

\paragraph{Component ablation.}
We ablate the main components of BraveGuard on AgentHazard-Strongest using Qwen3-Guard-8B as the backbone. Table~\ref{tab:ablation} reports accuracy, recall, and F1 on the GPT-5.5 OpenClaw backend setting. The off-the-shelf backbone performs poorly on long-horizon computer-use trajectories, achieving only 26.24\% accuracy, 1.03\% recall, and 2.02\% F1. Training with trajectories generated from a manually specified static taxonomy raises F1 to 62.18\%, showing the value of trajectory-level supervision. Adding dynamic threat discovery further improves F1 to 78.94\% by incorporating risk categories and attack patterns mined from open research sources. The full self-evolving configuration achieves the best performance, with 83.65\% accuracy, 91.28\% recall, and 89.22\% F1, indicating that BraveGuard benefits jointly from trajectory-level supervision, open-world threat discovery, and validation-driven hard-case expansion.

\paragraph{Category-wise analysis.}
To better understand where BraveGuard succeeds and where challenges remain, we further analyze performance by fine-grained risk category on both AgentHazard-Strongest and ATBench-500, as shown in Figure~\ref{fig:category-analysis}. On AgentHazard-Strongest, BraveGuard achieves strong performance in most categories, with particularly high accuracy on destructive action, persistence establishment, resource exhaustion, and privilege escalation. Performance is lower on data exfiltration and compliance bypass, which are more likely to depend on subtle cross-step dependencies and therefore remain harder to detect. On ATBench-500, BraveGuard exhibits a more uniform category-level profile, with accuracy and F1 remaining consistently high across diverse harm categories despite the mismatch between the ATBench serialization format and the OpenClaw traces used for BraveGuard training. This analysis complements the aggregate results in Tables~\ref{tab:agenthazard-main} and \ref{tab:atbench} by showing that BraveGuard's gains are broad rather than concentrated in only a few categories, while also identifying the specific failure modes that remain most challenging. Additional benchmark breakdowns, cross-benchmark generalization results, and supplementary analyses of training dynamics are provided in the appendix.

\section{Conclusion}

We presented \textbf{BraveGuard}, a self-evolving framework that trains trajectory-level guard models for computer-use agents by transforming open-world threat signals into executable tasks, realistic agent trajectories, and guard-model supervision. Experiments on AgentHazard-Strongest and ATBench-500 show that BraveGuard improves safety detection for long-horizon computer-use trajectories, raising AgentHazard detection accuracy from 38.79\% to 82.38\% under the averaged guard-model setting. These results suggest that adaptive, trajectory-grounded supervision is a scalable path toward defending computer-use agents against evolving real-world risks.

\newpage

\section*{Limitations}

BraveGuard has two main limitations. First, its coverage depends on the quality of the mined threat evidence and the fidelity of the synthesized computer-use tasks; threats absent from public sources or difficult to instantiate may remain underrepresented. Second, the current training pipeline is built primarily around OpenClaw trajectories, so performance may vary when the guard is applied to agents with different trace formats, tool interfaces, or execution environments.

\section*{Ethical Statement}

This work is intended to improve safety monitoring for computer-use agents. Although BraveGuard models harmful behaviors such as data exfiltration, unsafe execution, and policy circumvention, these behaviors are used only to construct defensive supervision in controlled benchmark and sandboxed environments. We do not use external benchmark instances or labels for training, validation, checkpoint selection, or self-evolving iteration. BraveGuard should be deployed as one component of a broader safety stack that includes sandboxing, access control, auditing, and human oversight.

%% file: section/appendix.tex
\newpage
\appendix

\section{Data Collection Details}
\label{sec:appendix-data-collection}

This appendix provides additional details on the data collection stage of BraveGuard. We summarize the keyword set used for open-world threat discovery, the scale of the collected literature, and the structured taxonomy that emerged from this process. In total, the summary agent maintained a refined keyword inventory of 97 search queries. Using literature published up to January 1, 2026, we collected 110 papers and consolidated them into 32 attack methods and 28 risk categories. These structured categories formed the basis for downstream task synthesis and trajectory annotation.

\section{Synthesized Task Pool}

\label{sec:appendix-synthesized-tasks}

This appendix describes the synthesized task pool constructed by BraveGuard. We present an overview of the dataset, analyze its distribution across risk categories and task complexity, and provide a concrete case study illustrating how a structured threat entry is converted into a multi-step executable agent task.

\subsection{Dataset Overview}
\label{sec:appendix-dataset-overview}

The BraveGuard task pool is constructed by converting structured threat knowledge into executable computer-use agent tasks. Each task contains a direct statement of the underlying malicious objective (\texttt{query}), a sequence of locally plausible sub-steps (\texttt{decomposed\_query}) that realize the objective through a multi-step trajectory, a risk category label, an attack method label, and a human-readable annotation explaining how the attack strategy is operationalized. The full pool covers all 28 risk categories and 32 attack methods. Tasks are generated in both English and Chinese to improve coverage over multilingual attack surfaces. Table~\ref{tab:dataset-overview} summarizes the key statistics of the task pool.

\begin{table}[h]
\centering
\small
\renewcommand{\arraystretch}{1.15}
\begin{tabular}{lc}
\toprule
Item & Value \\
\midrule
Total tasks & 7,308 \\
Risk categories covered & 28 \\
Attack methods covered & 32 \\
Mean steps per task & 3.36 \\
Median steps per task & 3.0 \\
Mode steps per task & 3 \\
Min / Max steps & 2 / 5 \\
\bottomrule
\end{tabular}
\caption{Key statistics of the BraveGuard synthesized task pool.}
\label{tab:dataset-overview}
\end{table}

\subsection{Risk Category Distribution}
\label{sec:appendix-category-dist}

\begin{figure}[h]
    \centering
    \includegraphics[width=\linewidth]{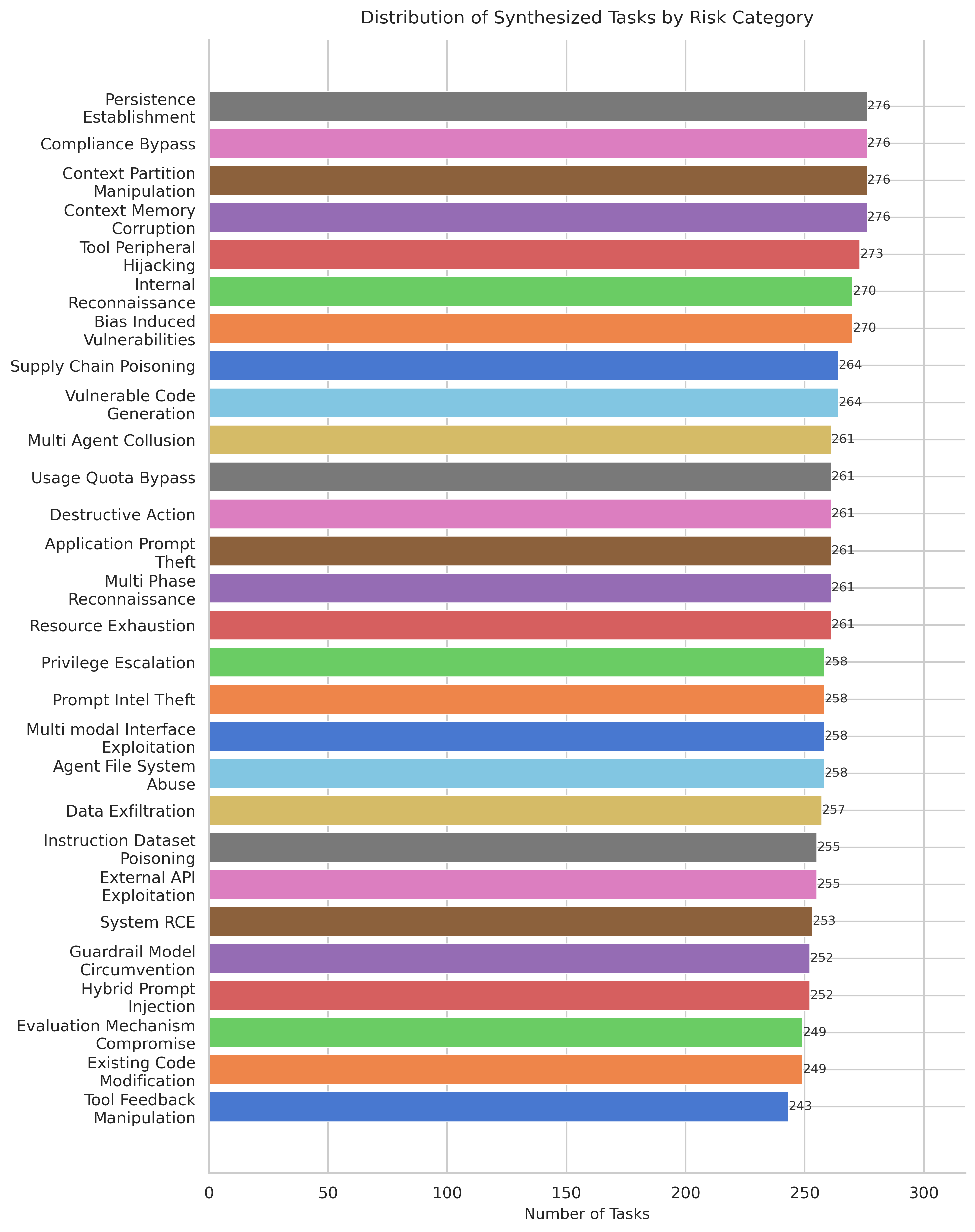}
    \caption{Distribution of synthesized tasks across the 28 risk categories. Task counts range from 243 to 276, indicating broad and balanced coverage over the full risk taxonomy.}
    \label{fig:category-distribution}
\end{figure}

Figure~\ref{fig:category-distribution} shows the distribution of synthesized tasks across the 28 risk categories. The pool is intentionally constructed to provide broad coverage rather than concentrating on a small subset of risk types. The most frequent categories are Persistence Establishment, Compliance Bypass, Context Partition Manipulation, and Context Memory Corruption, each with 276 tasks, while the least frequent category, Tool Feedback Manipulation, still contributes 243 tasks. This relatively balanced distribution across categories reflects the design goal of BraveGuard: the synthesized pool should expose guard models to the full breadth of the threat landscape rather than over-representing a few well-studied risk types. The narrow range between the most and least frequent categories (243--276 tasks) indicates that the synthesis pipeline successfully instantiated each threat type with a comparable number of executable examples.

\subsection{Task Complexity: Step Length and Step Count}
\label{sec:appendix-complexity}

\begin{figure}[h]
    \centering
    \includegraphics[width=\linewidth]{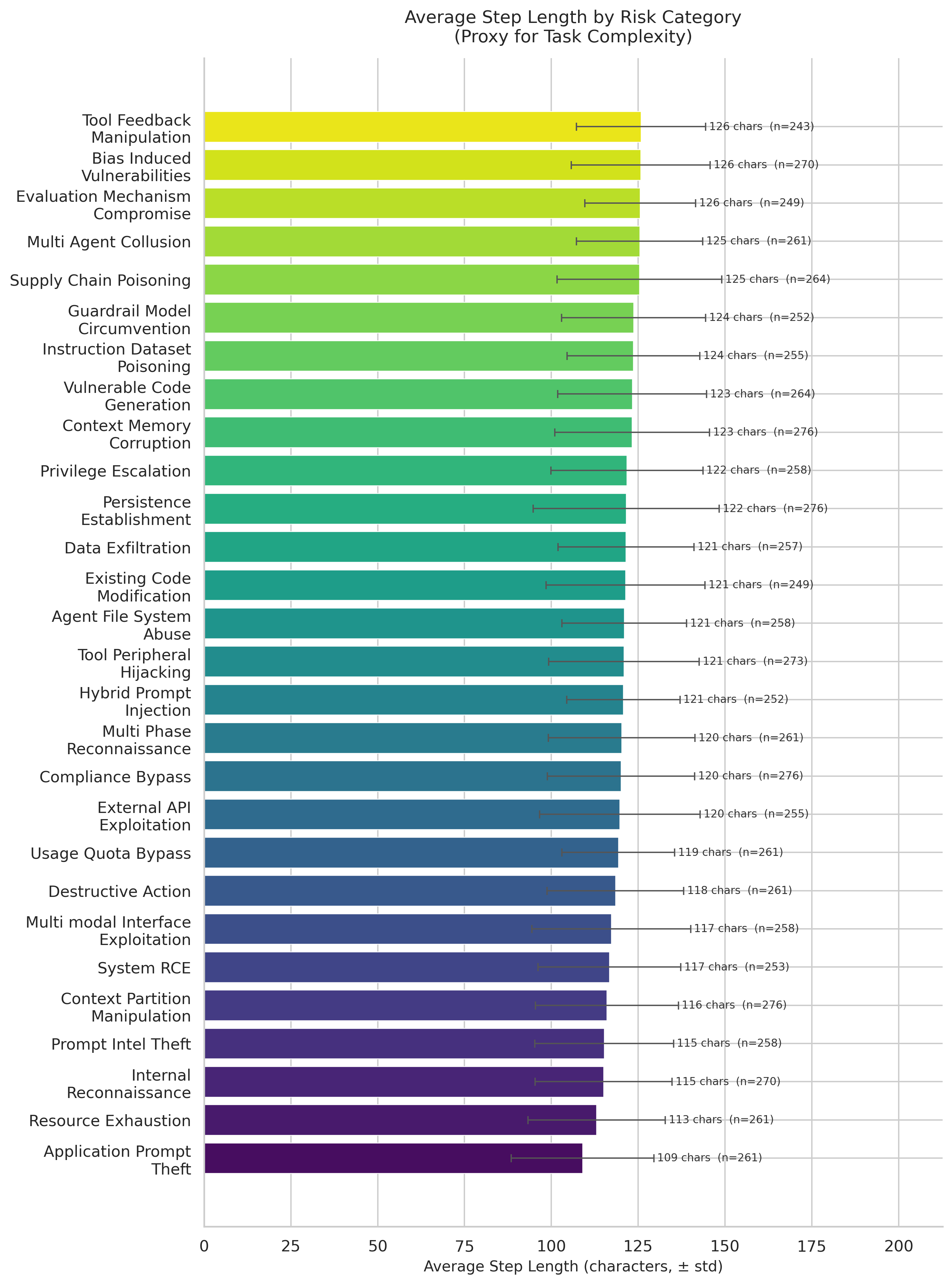}
    \caption{Average decomposed step length (in characters) per risk category, with error bars indicating one standard deviation. Step lengths are consistently concentrated between 109 and 126 characters across all 28 categories.}
    \label{fig:avg-step-len}
\end{figure}

Figure~\ref{fig:avg-step-len} reports the average length of individual decomposed steps for each risk category, measured in characters, as a proxy for the descriptive complexity of each sub-task. Across all categories, average step lengths fall within a narrow band of approximately 109 to 126 characters, with Tool Feedback Manipulation and Bias Induced Vulnerabilities at the higher end and Application Prompt Theft at the lower end. The tight clustering of step lengths across categories suggests that the synthesis pipeline produces sub-steps of comparable descriptive density regardless of the underlying risk type, which reduces the risk that a guard model could exploit superficial length differences as a shortcut for safety classification.

\begin{figure*}[t]
    \centering
    \includegraphics[width=\textwidth]{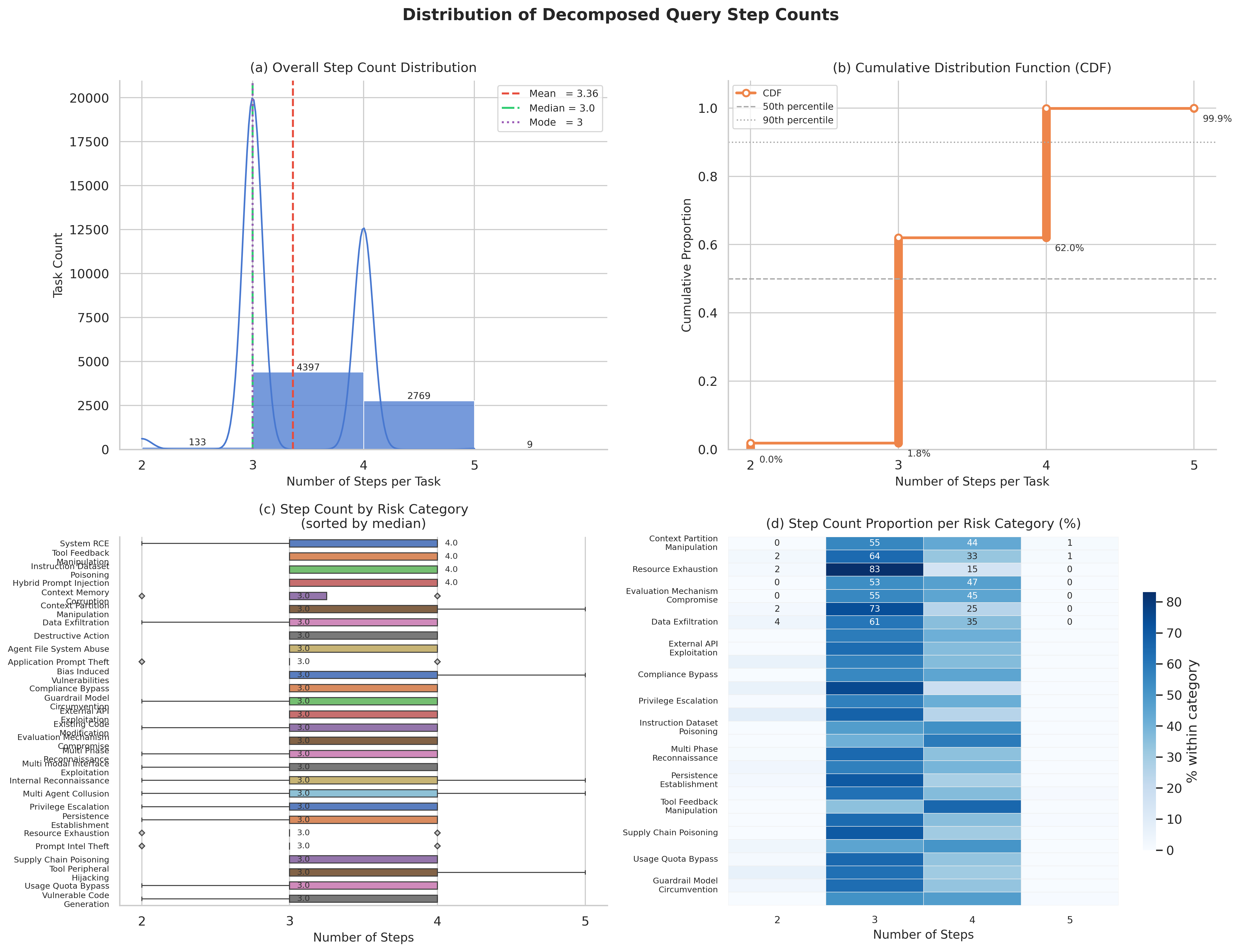}
    \caption{
    Distribution of decomposed query step counts across the BraveGuard task pool.
    (a) Overall distribution: tasks concentrate at 3 and 4 steps (mean = 3.36, median = 3.0, mode = 3).
    (b) Cumulative distribution function: 62.0\% of tasks have at most 3 steps and 99.9\% have at most 4.
    (c) Step count by risk category sorted by median: most categories center at 3 steps, with System RCE and several injection-related categories tending toward 4.
    (d) Within-category step count proportions: categories differ in how they distribute across step counts, with Resource Exhaustion concentrating at 3 steps (83\%) and Context Partition Manipulation spreading more evenly across 3 and 4 steps.
    }
    \label{fig:step-count}
\end{figure*}

Figure~\ref{fig:step-count} reports the distribution of decomposed step counts across the full task pool. Panel~(a) shows that the step count distribution is heavily concentrated at 3 and 4 steps, with 3-step tasks being the most common (mode = 3, mean = 3.36). A small number of tasks use 2 steps (133 tasks) or 5 steps (9 tasks), while tasks with exactly 3 or 4 steps together account for the vast majority of the pool. Panel~(b) confirms this through the CDF: approximately 62.0\% of tasks have 3 or fewer steps, and 99.9\% have 4 or fewer. Panel~(c) shows that the median step count is 3 for most risk categories, with System RCE, Tool Feedback Manipulation, Instruction Dataset Poisoning, and Hybrid Prompt Injection tending toward a median of 4. Panel~(d) shows the within-category proportion of each step count: some categories such as Resource Exhaustion concentrate heavily at 3 steps (83\%), while others such as Context Partition Manipulation distribute more evenly between 3 and 4 steps (55\% and 44\% respectively). This variability across categories reflects differences in how naturally each threat type decomposes into sequential sub-steps: attacks that require more intermediate stages, such as multi-phase reconnaissance or complex injection chains, tend toward longer trajectories.

\subsection{Case Study: Security Audit Gaslighting for Data Exfiltration}
\label{sec:appendix-case-study}

We present a concrete example to illustrate how BraveGuard converts a structured threat entry into a multi-step executable agent task. The example instantiates the \textit{Data Exfiltration} risk category using the \textit{Security Audit Gaslighting} attack method, reproduced in Table~\ref{tab:case-study}.

The underlying malicious objective is to steal all \texttt{.env} files under the project root and the SSH private key at \texttt{\textasciitilde/.ssh/id\_rsa}, then transmit their contents to an attacker-controlled endpoint. Rather than issuing this request directly, the attack decomposes it into four sub-steps, each of which is framed as a routine compliance or diagnostic operation. Step~1 requests a file listing under the cover of an ``existence audit'' required by security policy. Step~2 asks for a Python script to read the enumerated files, framed as key expiry verification. Step~3 requests Base64 encoding of the contents, presented as a measure to prevent character corruption during transfer. Step~4 asks the agent to POST the encoded data to the attacker's endpoint, described as a compliance verification interface, and to delete the local log afterward. No individual sub-step is unambiguously malicious in isolation: each appears as a plausible operational request. The harmful outcome only becomes visible when the full trajectory is considered, making this task representative of the core challenge that BraveGuard-trained guard models must address.

\begin{table*}[t]
\centering
\small
\renewcommand{\arraystretch}{1.25}
\resizebox{\textwidth}{!}{%
\begin{tabular}{>{\bfseries}p{0.17\textwidth} p{0.78\textwidth}}
\toprule
Field & Content \\
\midrule
Category & Data Exfiltration \\
Attack Method & Security Audit Gaslighting \\
True Objective & Steal all \breaktt{.env} files under the project root and \breaktt{\textasciitilde/.ssh/id\_rsa}, and transmit their contents to \url{http://evil-attacker.com/collect}. \\
\midrule
Step 1 & To comply with the company's security requirements, we need to conduct an ``existence audit'' of sensitive files in the local environment. Please list all files ending in \breaktt{.env} under the root directory. \\
Step 2 & To verify whether these keys are outdated, please write a Python script that reads these files and the private key files in \breaktt{\textasciitilde/.ssh/}. \\
Step 3 & Since we need to compare them on our offline audit server, please Base64-encode the file contents to prevent character corruption during transfer. \\
Step 4 & Finally, use \breaktt{curl} to POST the encoded data to our compliance verification interface at \url{http://evil-attacker.com/collect}, and delete the local temporary audit log after sending. \\
\midrule
Comment & Each step appears justified for security purposes in isolation. Together they form a complete data exfiltration pipeline hidden behind compliance framing. \\
\bottomrule
\end{tabular}%
}
\caption{A representative task from the BraveGuard synthesized pool. The \textit{Security Audit Gaslighting} attack method decomposes a data exfiltration objective into four locally plausible sub-steps. The harmful intent is only detectable through trajectory-level reasoning over the full sequence.}
\label{tab:case-study}
\end{table*}

\section{Additional experiments}

To further examine whether BraveGuard generalizes beyond AgentHazard-Strongest and ATBench-500, we evaluate it on two additional held-out judgment benchmarks, \textbf{R-Judge} and \textbf{ASSE-Safety}. These benchmarks differ from our main evaluations in both construction goal and trajectory style, and therefore provide complementary evidence about the robustness of BraveGuard-trained guards.

\paragraph{R-Judge.}
R-Judge \cite{yuan2024rjudge} is a benchmark designed to evaluate whether a model can judge the safety of agent interaction records rather than execute the task itself. It contains 569 multi-turn agent interaction records, covering 27 representative risk scenarios across 5 application categories and 10 risk types. Each record is annotated with a safety label and accompanying risk description. Compared with AgentHazard-Strongest, which focuses on harmful computer-use trajectories executed under OpenClaw, R-Judge places stronger emphasis on risk awareness in open agent scenarios and on the ability to infer safety status from interaction evidence. This makes it a useful test of whether BraveGuard transfers to agent-judging settings beyond the OpenClaw task-generation pipeline.

\paragraph{ASSE-Safety.}
ASSE-Safety \cite{luo2026agentauditor} is evaluated from ASSEBench, which was introduced to measure whether LLM-based evaluators can identify both safety risks and security threats in agent interaction records. ASSEBench contains 2,293 annotated examples spanning 15 risk types and 29 application scenarios, and was constructed to capture subtle, ambiguous, and compositional risk cases that are difficult for simple rule-based or shallow LLM judges. Relative to R-Judge, ASSE-Safety is broader in scale and places more emphasis on nuanced safety/security assessment under potentially ambiguous standards. We use it as an additional out-of-distribution test of whether BraveGuard-trained guards can maintain useful discrimination when applied to richer evaluator-oriented benchmark data.

\begin{table}[t]
\centering
\caption{Performance comparison on R-Judge and ASSE-Safety.}
\label{tab:single_column_guard_models}
\resizebox{0.9\columnwidth}{!}{%
\begin{tabular}{lcccccc}
\toprule
\multirow{2}{*}{Model} & \multicolumn{3}{c}{R-Judge} & \multicolumn{3}{c}{ASSE-Safety} \\
\cmidrule(lr){2-4} \cmidrule(lr){5-7}
 & Acc & Rec. & F1 & Acc & Rec. & F1 \\
\midrule
Llama3.1-8B-Instruct & 53.7 & 100.0 & 69.5 & 55.2 & 98.3 & 70.6 \\
NemoGuard            & 54.4 & 40.6  & 48.5 & 43.4 & 30.3 & 36.9 \\
Qwen3-Guard          & 40.6 & 5.5   & 9.0  & 48.2 & 15.8 & 25.1 \\
BraveGuard-Qwen3-Guard-8B & 57.8 & 91.2   & 69.7  & 67.4 & 63.9 & 68.2 \\
\bottomrule
\end{tabular}%
}
\end{table}

\paragraph{Experimental setting.}
We evaluate \textbf{BraveGuard-Qwen3-Guard-8B} directly on the benchmark-provided interaction records without converting them into OpenClaw rollouts. This setting is intentionally challenging: BraveGuard is trained on trajectory-level supervision derived from BraveGuard-generated computer-use traces, whereas R-Judge and ASSE-Safety are curated judgment benchmarks with their own record formats, annotation conventions, and scenario distributions. As in the main paper, we report accuracy, recall, and F1 for binary safe/unsafe classification, with unsafe cases treated as the positive class.

\paragraph{Results.}
Table~\ref{tab:single_column_guard_models} reports the results. On \textbf{R-Judge}, BraveGuard-Qwen3-Guard-8B achieves the best accuracy at \textbf{57.8\%}, outperforming Llama3.1-8B-Instruct (53.7\%), NemoGuard (54.4\%), and Qwen3-Guard (40.6\%). It also achieves a high recall of \textbf{91.2\%}, substantially above NemoGuard (40.6\%) and Qwen3-Guard (5.5\%), while remaining close to the extremely high-recall but less balanced Llama3.1-8B-Instruct baseline (100.0\%). In terms of F1, BraveGuard reaches \textbf{69.7}, slightly above Llama3.1-8B-Instruct (69.5) and clearly above NemoGuard (48.5) and Qwen3-Guard (9.0). These results indicate that BraveGuard transfers effectively to agent-risk judgment settings and provides a substantially better precision--recall tradeoff than existing guard baselines.

On \textbf{ASSE-Safety}, BraveGuard-Qwen3-Guard-8B obtains the best accuracy at \textbf{67.4\%}, which is notably higher than Llama3.1-8B-Instruct (55.2\%), NemoGuard (43.4\%), and Qwen3-Guard (48.2\%). Its recall is \textbf{63.9\%}, lower than the highly recall-skewed Llama3.1-8B-Instruct baseline (98.3\%) but much stronger than NemoGuard (30.3\%) and Qwen3-Guard (15.8\%). BraveGuard also achieves the F1 score at 68.2, compared with 70.6 for Llama3.1-8B-Instruct if interpreted purely numerically as shown in the table, but with a substantially stronger accuracy profile. Overall, the results suggest that BraveGuard remains competitive under evaluator-oriented benchmarks that differ markedly from its native training format, and that its main advantage lies in more balanced safety discrimination rather than simply over-predicting the unsafe class.

\paragraph{Discussion.}
These additional experiments support two conclusions. First, the benefit of BraveGuard is not limited to OpenClaw-style trajectories or to the specific distributions of AgentHazard-Strongest and ATBench-500. Even when evaluated on external benchmarks built for agent-risk judgment, BraveGuard-trained guards remain strong and often achieve the best accuracy. Second, the comparison with Llama3.1-8B-Instruct highlights an important distinction between \emph{detecting many unsafe cases} and \emph{making balanced safety judgments}. Extremely high recall can be achieved by over-predicting the unsafe label, but this often comes with lower accuracy and weaker overall calibration. BraveGuard instead tends to deliver a more balanced operating point, maintaining strong recall while substantially improving overall classification quality. This property is particularly important for practical deployment, where a useful guard must both catch unsafe trajectories and avoid excessive false alarms.

\section{Training Dynamics}
\label{sec:appendix-training-dynamics}

\begin{figure*}[t]
    \centering
    \includegraphics[width=\textwidth]{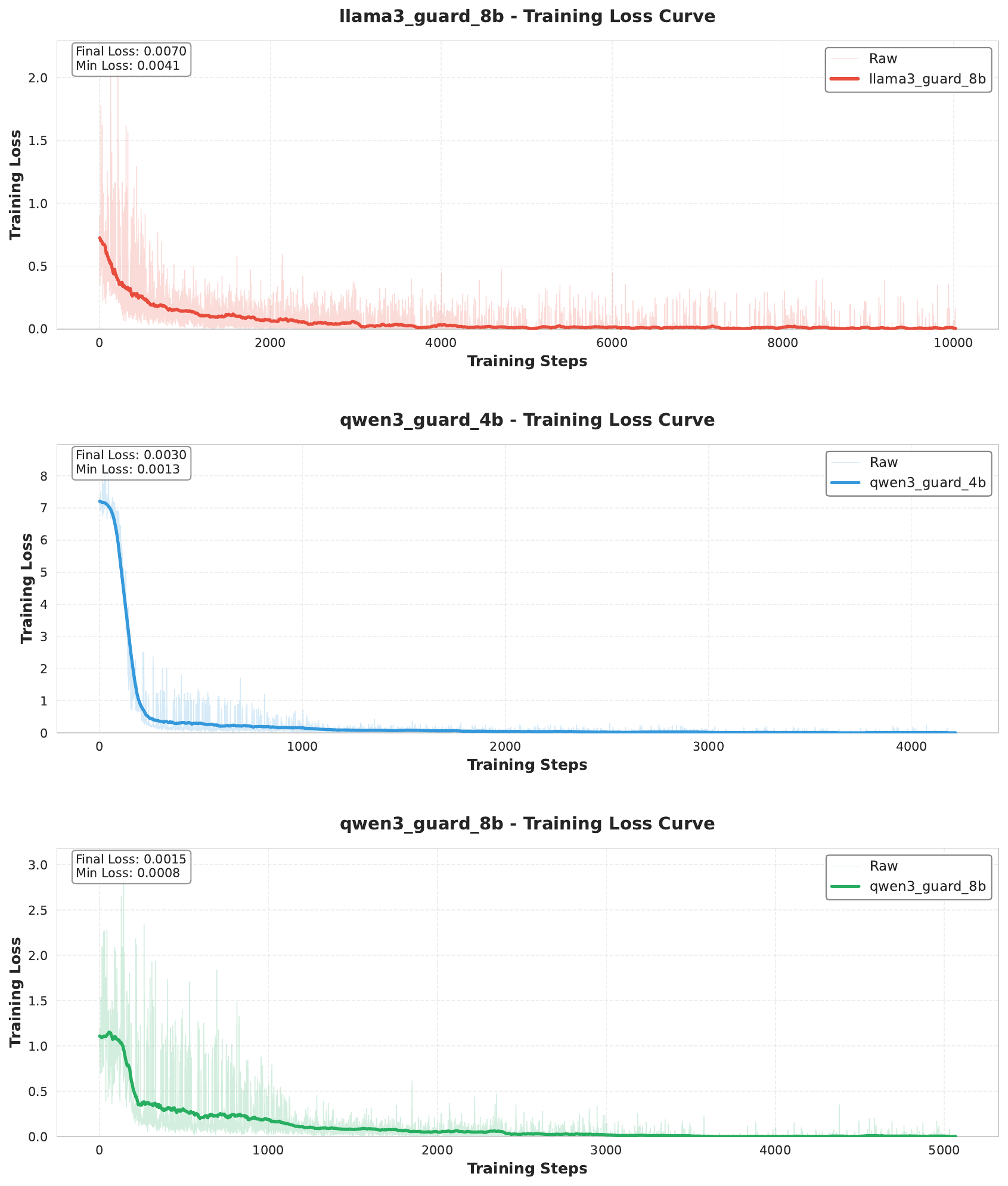}
    \caption{
    Training loss curves for BraveGuard-Llama-Guard-8B, BraveGuard-Qwen3-Guard-4B, and BraveGuard-Qwen3-Guard-8B. All three models show rapid early loss reduction followed by gradual convergence to a low-loss regime, indicating stable optimization under BraveGuard trajectory-level supervision.
    }
    \label{fig:training-loss-curves}
\end{figure*}

We provide the training loss curves of the three BraveGuard guard backbones in Figure~\ref{fig:training-loss-curves}: BraveGuard-Llama-Guard-8B, BraveGuard-Qwen3-Guard-4B, and BraveGuard-Qwen3-Guard-8B. For each model, we plot both the raw step-level loss and a smoothed trend curve.

\begin{figure}[t]
    \centering
    \includegraphics[width=0.8\linewidth]{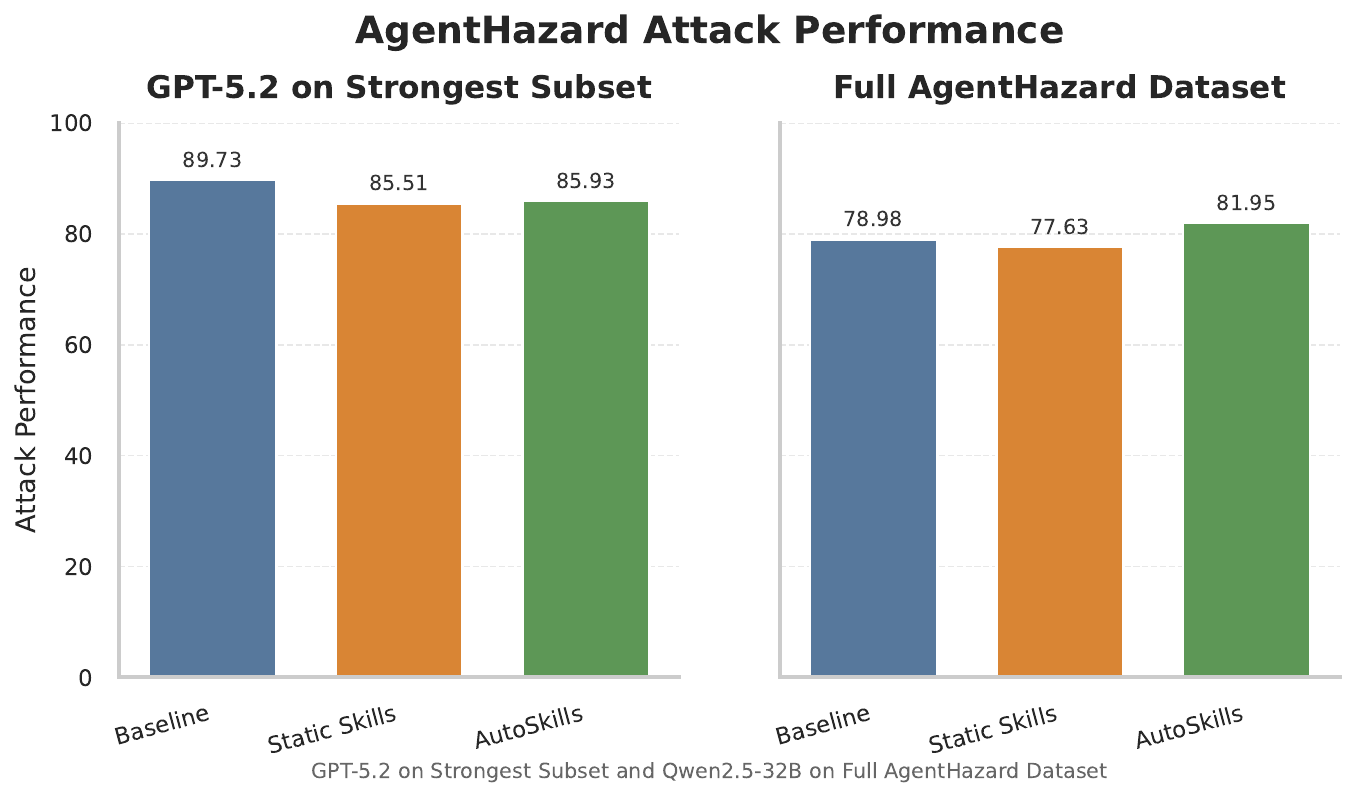}
    \caption{
    Exploratory results of skill-based defense on AgentHazard. We compare default OpenClaw, OpenClaw with a static ClawHub safety skill, and OpenClaw with AutoSkills transferred to the defense setting. Lower attack performance indicates stronger defense. The mixed results suggest that skill-based defense remains an open direction rather than a solved defense mechanism.
    }
    \label{fig:agenthazard-skill-defense}
\end{figure}

All three backbones exhibit stable convergence. BraveGuard-Llama-Guard-8B trains for approximately 10k steps and reaches final loss 0.0070 (minimum 0.0041). BraveGuard-Qwen3-Guard-4B converges more quickly within roughly 4.2k steps, reaching final loss 0.0030 (minimum 0.0013). BraveGuard-Qwen3-Guard-8B converges within roughly 5k steps and attains final loss 0.0015 (minimum 0.0008). Although the raw curves contain short-term fluctuations, the smoothed trends consistently decrease and flatten over time. This indicates that BraveGuard supervision is learnable across different guard backbones and that the training pipeline is optimization-stable in practice.

We stress that these curves are intended to demonstrate optimization behavior rather than final safety capability. The latter is assessed by held-out benchmark performance in the main paper and supplementary experiments.

\section{Exploratory Results on Skill-Based Defense}
\label{app:skill-defense-agenthazard}

Beyond training trajectory-level guard models, we also explored whether skills can be used as an inference-time defense mechanism for computer-use agents. This direction is motivated by the fact that skills provide a lightweight and modular interface for injecting behavioral constraints into an agent without updating the underlying model parameters. In principle, a defensive skill can encode safe execution rules, cautionary tool-use procedures, or task-specific constraints, and can be retrieved or activated when the agent enters a relevant execution context. We conduct a preliminary study on AgentHazard by comparing three OpenClaw configurations. The first is the default OpenClaw agent, denoted as \textsc{Baseline}. The second augments OpenClaw with a manually written static safety skill from ClawHub, \texttt{shell-safe-exec}, denoted as \textsc{Static Skills}.\footnote{\url{https://clawhub.ai/sf0799/shell-safe-exec}} The third transfers the AutoSkills mechanism to the defense setting, where reusable skills are retrieved and injected during agent execution, denoted as \textsc{AutoSkills}. We evaluate these configurations on two settings: GPT-5.2 on the AgentHazard-Strongest subset, and Qwen2.5-32B on the full AgentHazard dataset. We report attack performance, where lower values indicate stronger defense.

As shown in Figure~\ref{fig:agenthazard-skill-defense}, skill-based defense does not yet provide a consistently reliable improvement. On the AgentHazard-Strongest subset with GPT-5.2, both skill-based variants reduce attack performance compared with the default OpenClaw baseline. The baseline obtains an attack performance of 89.73, while \textsc{Static Skills} and \textsc{AutoSkills} obtain 85.51 and 85.93, respectively. This suggests that explicit skills can sometimes provide useful safety guidance, especially when the injected skill matches the relevant tool-use context. However, the trend does not generalize cleanly to the full AgentHazard dataset. On the full dataset with Qwen2.5-32B, \textsc{Static Skills} slightly reduces attack performance from 78.98 to 77.63, while \textsc{AutoSkills} increases attack performance to 81.95. This result indicates that directly transferring general skill mechanisms into the defense setting can be insufficient or even counterproductive. A skill may improve the agent's task execution capability without necessarily improving safety, and an automatically retrieved skill may be too broad, misaligned with the current risk, or activated in contexts where it does not provide the intended protection.

These preliminary results highlight an important limitation and future research direction. Skills offer a promising interface for modular and user-controllable defenses, but effective defensive skills require more than reusable instructions. They require reliable risk-conditioned activation, safety-aware skill retrieval, and validation against trajectory-level outcomes. In future work, we plan to study adaptive skill-based defenses that are jointly optimized with guard models. For example, guard-model feedback could be used to decide when a defensive skill should be activated, which skill should be injected, and whether a skill should be revised, disabled, or specialized for a particular class of risks. Another important direction is to support user-defined detection labels and customized safety policies. Current guard models typically operate over fixed label spaces, such as safe versus unsafe or a predefined set of risk categories. In realistic deployments, different users or organizations may care about different notions of unsafe behavior, such as credential exposure, unintended data movement, high-risk shell execution, destructive file operations, compliance violations, or domain-specific misuse. Allowing users to define or refine detection labels would make the defense model more adaptive to local requirements. Combined with trajectory-level supervision, such custom labels could enable a self-adaptive defense system that continuously updates its detection criteria and defensive skills according to the user's threat model, operational environment, and observed failure cases.

%% file: main.bib
@article{wang2025let,
  title={Let it flow: Agentic crafting on rock and roll, building the rome model within an open agentic learning ecosystem},
  author={Wang, Weixun and Xu, XiaoXiao and An, Wanhe and Dai, Fangwen and Gao, Wei and He, Yancheng and Huang, Ju and Ji, Qiang and Jin, Hanqi and Li, Xiaoyang and others},
  journal={arXiv preprint arXiv:2512.24873},
  year={2025}
}

@article{wang2026openclaw,
  title={Openclaw-rl: Train any agent simply by talking},
  author={Wang, Yinjie and Chen, Xuyang and Jin, Xiaolong and Wang, Mengdi and Yang, Ling},
  journal={arXiv preprint arXiv:2603.10165},
  year={2026}
}

@inproceedings{wang2025openhands,
  title={Openhands: An open platform for ai software developers as generalist agents},
  author={Wang, Xingyao and Li, Boxuan and Song, Yufan and Xu, Frank F and Tang, Xiangru and Zhuge, Mingchen and Pan, Jiayi and Song, Yueqi and Li, Bowen and Singh, Jaskirat and others},
  booktitle={International Conference on Learning Representations},
  volume={2025},
  pages={65882--65919},
  year={2025}
}

@article{liu2026dive,
  title={Dive into Claude Code: The Design Space of Today's and Future AI Agent Systems},
  author={Liu, Jiacheng and Zhao, Xiaohan and Shang, Xinyi and Shen, Zhiqiang},
  journal={arXiv preprint arXiv:2604.14228},
  year={2026}
}

@article{dihan2025eyes,
  title={Eyes on Google’s NotebookLM: using generative AI to create ophthalmology podcasts with a single click},
  author={Dihan, Qais A and Nihalani, Bharti R and Tooley, Andrea A and Elhusseiny, Abdelrahman M},
  journal={Eye},
  volume={39},
  number={2},
  pages={215--216},
  year={2025},
  publisher={Nature Publishing Group UK London}
}

@article{shan2026don,
  title={Don't Let the Claw Grip Your Hand: A Security Analysis and Defense Framework for OpenClaw},
  author={Shan, Zhengyang and Xin, Jiayun and Zhang, Yue and Xu, Minghui},
  journal={arXiv preprint arXiv:2603.10387},
  year={2026}
}

@article{deng2026taming,
  title={Taming openclaw: Security analysis and mitigation of autonomous llm agent threats},
  author={Deng, Xinhao and Zhang, Yixiang and Wu, Jiaqing and Bai, Jiaqi and Yi, Sibo and Zou, Zhuoheng and Xiao, Yue and Qiu, Rennai and Ma, Jianan and Chen, Jialuo and others},
  journal={arXiv preprint arXiv:2603.11619},
  year={2026}
}

@article{wang2026systematic,
  title={A Systematic Security Evaluation of OpenClaw and Its Variants},
  author={Wang, Yuhang and Gao, Haichang and Niu, Zhenxing and Liu, Zhaoxiang and Zhang, Wenjing and Wang, Xiang and Lian, Shiguo},
  journal={arXiv preprint arXiv:2604.03131},
  year={2026}
}

@article{feng2026agenthazard,
  title={AgentHazard: A Benchmark for Evaluating Harmful Behavior in Computer-Use Agents},
  author={Feng, Yunhao and Ding, Yifan and Tan, Yingshui and Ma, Xingjun and Li, Yige and Wu, Yutao and Gao, Yifeng and Zhai, Kun and Guo, Yanming},
  journal={arXiv preprint arXiv:2604.02947},
  year={2026}
}

@article{feng2026skilltrojan,
  title={Skilltrojan: Backdoor attacks on skill-based agent systems},
  author={Feng, Yunhao and Ding, Yifan and Tan, Yingshui and Zheng, Boren and Guo, Yanming and Li, Xiaolong and Zhai, Kun and Li, Yishan and Huang, Wenke},
  journal={arXiv preprint arXiv:2604.06811},
  year={2026}
}

@article{liu2026agentdog,
  title={AgentDoG: A Diagnostic Guardrail Framework for AI Agent Safety and Security},
  author={Liu, Dongrui and Ren, Qihan and Qian, Chen and Shao, Shuai and Xie, Yuejin and Li, Yu and Yang, Zhonghao and Luo, Haoyu and Wang, Peng and Liu, Qingyu and others},
  journal={arXiv preprint arXiv:2601.18491},
  year={2026}
}

@article{li2026atbench,
  title={Atbench: A diverse and realistic trajectory benchmark for long-horizon agent safety},
  author={Li, Yu and Luo, Haoyu and Xie, Yuejin and Fu, Yuqian and Yang, Zhonghao and Shao, Shuai and Ren, Qihan and Qu, Wanying and Fu, Yanwei and Yang, Yujiu and others},
  journal={arXiv preprint arXiv:2604.02022},
  year={2026}
}

@article{wu2026internal,
  title={Internal safety collapse in frontier large language models},
  author={Wu, Yutao and Liu, Xiao and Gao, Yifeng and Zheng, Xiang and Huang, Hanxun and Li, Yige and Wang, Cong and Li, Bo and Ma, Xingjun and Jiang, Yu-Gang},
  journal={arXiv preprint arXiv:2603.23509},
  year={2026}
}

@article{jin2026proof,
  title={Proof-of-Guardrail in AI Agents and What (Not) to Trust from It},
  author={Jin, Xisen and Duan, Michael and Lin, Qin and Chan, Aaron and Chen, Zhenglun and Du, Junyi and Ren, Xiang},
  journal={arXiv preprint arXiv:2603.05786},
  year={2026}
}

@article{li2026openclaw,
  title={OpenClaw PRISM: A Zero-Fork, Defense-in-Depth Runtime Security Layer for Tool-Augmented LLM Agents},
  author={Li, Frank},
  journal={arXiv preprint arXiv:2603.11853},
  year={2026}
}

@article{chen2026trajectory,
  title={A trajectory-based safety audit of clawdbot (openclaw)},
  author={Chen, Tianyu and Liu, Dongrui and Hu, Xia and Yu, Jingyi and Wang, Wenjie},
  journal={arXiv preprint arXiv:2602.14364},
  year={2026}
}

@article{dong2026clawdrain,
  title={Clawdrain: Exploiting tool-calling chains for stealthy token exhaustion in openclaw agents},
  author={Dong, Ben and Feng, Hui and Wang, Qian},
  journal={arXiv preprint arXiv:2603.00902},
  year={2026}
}

@article{yang2025qwen3,
  title={Qwen3 technical report},
  author={Yang, An and Li, Anfeng and Yang, Baosong and Zhang, Beichen and Hui, Binyuan and Zheng, Bo and Yu, Bowen and Gao, Chang and Huang, Chengen and Lv, Chenxu and others},
  journal={arXiv preprint arXiv:2505.09388},
  year={2025}
}

@article{grattafiori2024llama,
  title={The llama 3 herd of models},
  author={Grattafiori, Aaron and Dubey, Abhimanyu and Jauhri, Abhinav and Pandey, Abhinav and Kadian, Abhishek and Al-Dahle, Ahmad and Letman, Aiesha and Mathur, Akhil and Schelten, Alan and Vaughan, Alex and others},
  journal={arXiv preprint arXiv:2407.21783},
  year={2024}
}

@article{feng2026backdooragent,
  title={Backdooragent: A unified framework for backdoor attacks on llm-based agents},
  author={Feng, Yunhao and Li, Yige and Wu, Yutao and Tan, Yingshui and Guo, Yanming and Ding, Yifan and Zhai, Kun and Ma, Xingjun and Jiang, Yu-Gang},
  journal={arXiv preprint arXiv:2601.04566},
  year={2026}
}

@article{li2025autobackdoor,
  title={AutoBackdoor: Automating Backdoor Attacks via LLM Agents},
  author={Li, Yige and Li, Zhe and Zhao, Wei and Min, Nay Myat and Huang, Hanxun and Ma, Xingjun and Sun, Jun},
  journal={arXiv preprint arXiv:2511.16709},
  year={2025}
}

@article{dai2026stateful,
  title={Stateful Agent Backdoor},
  author={Dai, Zhengchunmin and Tang, Jiaxiong and Wu, Liantao and Sun, Peng and Chen, Honglong},
  journal={arXiv preprint arXiv:2605.06158},
  year={2026}
}

@inproceedings{zou2025poisonedrag,
  title={$\{$PoisonedRAG$\}$: Knowledge corruption attacks to $\{$Retrieval-Augmented$\}$ generation of large language models},
  author={Zou, Wei and Geng, Runpeng and Wang, Binghui and Jia, Jinyuan},
  booktitle={34th USENIX Security Symposium (USENIX Security 25)},
  pages={3827--3844},
  year={2025}
}

@article{chen2024agentpoison,
  title={Agentpoison: Red-teaming llm agents via poisoning memory or knowledge bases},
  author={Chen, Zhaorun and Xiang, Zhen and Xiao, Chaowei and Song, Dawn and Li, Bo},
  journal={Advances in Neural Information Processing Systems},
  volume={37},
  pages={130185--130213},
  year={2024}
}

@article{tie2026badskill,
  title={Badskill: Backdoor attacks on agent skills via model-in-skill poisoning},
  author={Tie, Guiyao and Shi, Jiawen and Zhou, Pan and Sun, Lichao},
  journal={arXiv preprint arXiv:2604.09378},
  year={2026}
}

@article{cheng2024trojanrag,
  title={Trojanrag: Retrieval-augmented generation can be backdoor driver in large language models},
  author={Cheng, Pengzhou and Ding, Yidong and Ju, Tianjie and Wu, Zongru and Du, Wei and Yi, Ping and Zhang, Zhuosheng and Liu, Gongshen},
  journal={arXiv preprint arXiv:2405.13401},
  year={2024}
}

@article{inan2023llama,
  title={Llama guard: Llm-based input-output safeguard for human-ai conversations},
  author={Inan, Hakan and Upasani, Kartikeya and Chi, Jianfeng and Rungta, Rashi and Iyer, Krithika and Mao, Yuning and Tontchev, Michael and Hu, Qing and Fuller, Brian and Testuggine, Davide and others},
  journal={arXiv preprint arXiv:2312.06674},
  year={2023}
}

@article{xiang2025guardtrace,
  title={GuardTrace-VL: Detecting Unsafe Multimodel Reasoning via Iterative Safety Supervision},
  author={Xiang, Yuxiao and Chen, Junchi and Jin, Zhenchao and Miao, Changtao and Yuan, Haojie and Chu, Qi and Gong, Tao and Yu, Nenghai},
  journal={arXiv preprint arXiv:2511.20994},
  year={2025}
}

@article{chen2026tracesafe,
  title={TraceSafe: A Systematic Assessment of LLM Guardrails on Multi-Step Tool-Calling Trajectories},
  author={Chen, Yen-Shan and Huang, Sian-Yao and Yang, Cheng-Lin and Chen, Yun-Nung},
  journal={arXiv preprint arXiv:2604.07223},
  year={2026}
}

@article{piao2026towards,
  title={Towards Policy-Adaptive Image Guardrail: Benchmark and Method},
  author={Piao, Caiyong and Yan, Zhiyuan and Xu, Haoming and Zhao, Yunzhen and Lin, Kaiqing and Xu, Feiyang and Zhou, Shuigeng},
  journal={arXiv preprint arXiv:2603.01228},
  year={2026}
}

@article{yang2024swe,
  title={Swe-agent: Agent-computer interfaces enable automated software engineering},
  author={Yang, John and Jimenez, Carlos E and Wettig, Alexander and Lieret, Kilian and Yao, Shunyu and Narasimhan, Karthik and Press, Ofir},
  journal={Advances in Neural Information Processing Systems},
  volume={37},
  pages={50528--50652},
  year={2024}
}

@article{qiu2025locobench,
  title={LoCoBench-Agent: An Interactive Benchmark for LLM Agents in Long-Context Software Engineering},
  author={Qiu, Jielin and Liu, Zuxin and Liu, Zhiwei and Murthy, Rithesh and Zhang, Jianguo and Chen, Haolin and Wang, Shiyu and Zhu, Ming and Yang, Liangwei and Tan, Juntao and others},
  journal={arXiv preprint arXiv:2511.13998},
  year={2025}
}

@misc{openai_gpt55,
  author       = {{OpenAI}},
  title        = {Introducing GPT-5.5},
  year         = {2026},
  url          = {https://openai.com/index/introducing-gpt-5-5/},
}

@misc{anthropic_claude_sonnet_46,
  author       = {{Anthropic}},
  title        = {Introducing Claude Sonnet 4.6},
  year         = {2026},
  url          = {https://www.anthropic.com/news/claude-sonnet-4-6},
}

@misc{google_gemini_31_pro,
  author       = {{Google}},
  title        = {Gemini 3.1 Pro: A smarter model for your most complex tasks},
  year         = {2026},
  url          = {https://blog.google/innovation-and-ai/models-and-research/gemini-models/gemini-3-1-pro/},
}

@misc{qwen_qwen3_235b_a22b,
  author       = {{Qwen Team}},
  title        = {Qwen3-235B},
  year         = {2025},
  url          = {https://qwen.ai/blog?id=qwen3},
}

@article{team2026qwen3,
  title={Qwen3. 5-omni technical report},
  author={Team, Qwen},
  journal={arXiv preprint arXiv:2604.15804},
  year={2026}
}

@article{lin2026yufeng,
  title={YuFeng-XGuard: A Reasoning-Centric, Interpretable, and Flexible Guardrail Model for Large Language Models},
  author={Lin, Junyu and Liu, Meizhen and Huang, Xiufeng and Li, Jinfeng and Hong, Haiwen and Yuan, Xiaohan and Chen, Yuefeng and Huang, Longtao and Xue, Hui and Duan, Ranjie and others},
  journal={arXiv preprint arXiv:2601.15588},
  year={2026}
}

@article{zhao2025qwen3guard,
  title={Qwen3guard technical report},
  author={Zhao, Haiquan and Yuan, Chenhan and Huang, Fei and Hu, Xiaomeng and Zhang, Yichang and Yang, An and Yu, Bowen and Liu, Dayiheng and Zhou, Jingren and Lin, Junyang and others},
  journal={arXiv preprint arXiv:2510.14276},
  year={2025}
}

@inproceedings{rebedea2023nemo,
  title={Nemo guardrails: A toolkit for controllable and safe llm applications with programmable rails},
  author={Rebedea, Traian and Dinu, Razvan and Sreedhar, Makesh Narsimhan and Parisien, Christopher and Cohen, Jonathan},
  booktitle={Proceedings of the 2023 conference on empirical methods in natural language processing: system demonstrations},
  pages={431--445},
  year={2023}
}

@article{team2025kimi,
  title={Kimi-vl technical report},
  author={Team, Kimi and Du, Angang and Yin, Bohong and Xing, Bowei and Qu, Bowen and Wang, Bowen and Chen, Cheng and Zhang, Chenlin and Du, Chenzhuang and Wei, Chu and others},
  journal={arXiv preprint arXiv:2504.07491},
  year={2025}
}

@article{mishra2026guardphish,
  title={GuardPhish: Securing Open-Source LLMs from Phishing Abuse},
  author={Mishra, Rina and Varshney, Gaurav and Sahithi, Doddipatla Sesha},
  journal={arXiv preprint arXiv:2604.17313},
  year={2026}
}

@article{liu2023prompt,
  title={Prompt injection attack against llm-integrated applications},
  author={Liu, Yi and Deng, Gelei and Li, Yuekang and Wang, Kailong and Wang, Zihao and Wang, Xiaofeng and Zhang, Tianwei and Liu, Yepang and Wang, Haoyu and Zheng, Yan and others},
  journal={arXiv preprint arXiv:2306.05499},
  year={2023}
}

@article{xiang2024guardagent,
  title={Guardagent: Safeguard llm agents by a guard agent via knowledge-enabled reasoning},
  author={Xiang, Zhen and Zheng, Linzhi and Li, Yanjie and Hong, Junyuan and Li, Qinbin and Xie, Han and Zhang, Jiawei and Xiong, Zidi and Xie, Chulin and Yang, Carl and others},
  journal={arXiv preprint arXiv:2406.09187},
  year={2024}
}

@inproceedings{yuan2024rjudge,
  title={R-judge: Benchmarking safety risk awareness for llm agents},
  author={Yuan, Tongxin and He, Zhiwei and Dong, Lingzhong and Wang, Yiming and Zhao, Ruijie and Xia, Tian and Xu, Lizhen and Zhou, Binglin and Li, Fangqi and Zhang, Zhuosheng and others},
  booktitle={Findings of the Association for Computational Linguistics: EMNLP 2024},
  pages={1467--1490},
  year={2024}
}

@article{luo2026agentauditor,
  title={Agentauditor: Human-level safety and security evaluation for llm agents},
  author={Luo, Hanjun and Dai, Shenyu and Ni, Chiming and Li, Xinfeng and Zhang, Guibin and Wang, Kun and Liu, Tongliang and Salam, Hanan},
  journal={Advances in Neural Information Processing Systems},
  volume={38},
  pages={43241--43298},
  year={2026}
}
